\documentclass[a4paper,11pt]{article}
\pdfoutput=1 

\usepackage{jheppub} 

\usepackage[T1]{fontenc}

\usepackage{float}
\usepackage{amsmath}
\usepackage{amssymb}
\usepackage{bm}
\usepackage{pbox}
\usepackage{placeins}
\usepackage{graphicx}
\usepackage[T1]{fontenc}
\usepackage{footnote}
\usepackage{pdfpages}
\usepackage{hhline}
\usepackage{multirow}
\usepackage{multicol}
\usepackage{enumitem}
\usepackage{slashed}
\usepackage{orcidlink}
\usepackage{booktabs}
\usepackage{cancel}
\allowdisplaybreaks

\usepackage{tikz}
\usepackage[compat=1.0.0]{tikz-feynman}

\usetikzlibrary{patterns}

\newcommand{\beqa}{\begin{eqnarray}}
\newcommand{\eeqa}{\end{eqnarray}}
\def\hc {\text{h.c.}}
\newcommand{\nn}{\nonumber}
\def\r {\rightarrow}

\newcommand{\GeV}{\mathrm{GeV}}
\newcommand{\MeV}{\mathrm{MeV}}

\newcommand{\del}{\partial}
\newcommand{\beq}{\begin{equation}}
\newcommand{\eeq}{\end{equation}}

\newcommand{\wb}{W\text{-boson}}

\newcommand{\braket}[1]{\ensuremath{\left\langle #1 \right\rangle}}

\newcommand{\gf}{G_F}
\newcommand{\gfit}{\texttt{Gfitter}}
\newcommand{\dth}{\delta_{\rm th}}
\newcommand{\seff}{\sin^2\theta_{\rm{eff}}^\ell}
\newcommand{\cm}{\chi^2_{\rm min}}

\title{Precision Electroweak Constraints on\\ Neutrinophilic Scalars}
	 
\author[a]{Saeid Foroughi-Abari\,\orcidlink{0000-0002-9406-1896},}
\author[a]{Camilla Mupo\,\orcidlink{0009-0002-3102-2723},}
\author[a,b]{Drona Vatsyayan\,\orcidlink{0000-0002-6868-3237},}
\author[a]{and Yue Zhang\,\orcidlink{0000-0002-1984-7450}\,}

\affiliation[a]{Department of Physics, Carleton University, 1125 Colonel By Drive, Ottawa, ON K1S 5B6, Canada}

\affiliation[b]{Arthur B. McDonald Canadian Astroparticle Physics Research Institute, Queen's University, 64 Bader Lane, Kingston, ON K7L 3N6, Canada}

\emailAdd{saeidf@physics.carleton.ca}
\emailAdd{camillamupo@cmail.carleton.ca}
\emailAdd{drona@physics.carleton.ca}
\emailAdd{yzhang@physics.carleton.ca}

\abstract{Strong self-interaction among the active neutrinos mediated by a neutrinophilic scalar is a well-motivated target of particle physics and cosmological probes. In this article, we present precision electroweak constraints on models for neutrino self-interaction. We first work in the simplified model where the finite radiative corrections are obtained with the guidance of gauge invariance. These corrections are logarithmically enhanced for small mediator masses. We point out the importance of neutrino charged-current coupling correction and its impact on the $\Delta r$ parameter and Fermi constant measurements. This effect was overlooked previously and allows us to derive the leading constraint on the neutrinophilic couplings for mediator mass above a few hundred MeV. We investigate the robustness of the result in a concrete UV completion which further includes a TeV-scale $SU(2)_L$ triplet scalar and find the simplified model constraints continue to hold for a wide range of parameter space. We pin down moving parts in the UV complete model and conditions when the contributions from heavy particle loops are no longer negligible. Our results serve as a useful road map for future explorations of the self-interacting neutrino paradigm.
}

\makeatletter
\gdef\@fpheader{}
\makeatother

\begin{document}
\maketitle
\flushbottom

%%%%%%%%%%%%%%%%%%%%%%%%%%%
\section{Introduction}
\label{sec:intro}
%%%%%%%%%%%%%%%%%%%%%%%%%%%

The neutrino sector bears a strong potential toward the next discovery of physics beyond the Standard Model (SM).
Novel self-interaction among neutrinos is a well-motivated candidate of new physics for alleviating tensions in the cosmological data~\cite{Kreisch:2019yzn, Das:2020xke, He:2023oke, Camarena:2023cku, Das:2023npl, Camarena:2024zck, Pal:2024yom, Racco:2024lbu} and providing a viable production mechanism for sterile neutrino dark matter~\cite{DeGouvea:2019wpf, Kelly:2020pcy, Kelly:2020aks, An:2023mkf, Benso:2024qrg, Parashari:2026dxo}.
The strength of neutrino self-interaction is often parametrized using a dimensionful coupling $G_{\rm eff}$ (analogue of the Fermi constant) for low-energy scattering processes.
Owing to the elusive nature of neutrinos, $G_{\rm eff}$ is allowed to be much higher than the actual Fermi constant.
In terms of microscopic theories, a large $G_{\rm eff}$ is enabled by the exchange of a light mediator (which will be called $\phi$ hereafter) with mass well below the weak scale.
To avoid stringent constraints on the coupling to charged leptons (see {\it e.g.} \cite{Bauer:2018onh, Ferber:2023iso}), we consider $\phi$ to be neutrinophilic.
The case of $\phi$ being a vector boson has been considered in~\cite{Laha:2013xua, Kelly:2020pcy, Berbig:2020wve} and strong constraints are found for small $\phi$ masses due to enhanced longitudinal mode production.
There is much more room for strong neutrino self-interaction if $\phi$ is a scalar boson~\cite{Berryman:2022hds}.
At low energies, the interacting Lagrangian between $\phi$ and neutrino is given by
\begin{equation}\label{eq:lagphi}
    \mathcal{L} = \sum_{\alpha,\beta=e,\mu,\tau} \frac{\lambda_{\alpha\beta}}{2} \bar \nu_\alpha \mathbb{P}_R \nu^c_\beta \phi + {\rm h.c.} \,
\end{equation}
where $\nu_{\alpha,\beta}$ denote neutrinos contained in the SM lepton doublet and the superscript $^{c}$ means charge-conjugation.
In this work, we assume that $\phi$ plays the sole role of mediating neutrino self-interaction and, for simplicity, it does not develop a vacuum expectation value (VEV).~\footnote{If $\phi$ had a VEV, it would contribute to the Majorana mass of neutrinos. In that case, the mass and coupling ranges of $\phi$ would be further restricted by the observed neutrino mixings and mass differences.}
This can be made technically natural if $\phi$ is a complex scalar and assigned two units of lepton number (or $B-L$)~\cite{Berryman:2018ogk}.
For the model to be phenomenologically viable,~\footnote{{\it E.g.}, suppressing the contribution to $Z$-boson invisible width from the $Z\to\phi\phi^*\to \nu\nu\bar\nu\bar\nu$ decay channel, which will be discussed in detail in section~\ref{sec:UVcompletion}.} the light $\phi$ must be (approximately) a singlet under the SM gauge symmetry.
The Lagrangian Eq.~\eqref{eq:lagphi} can be obtained from a gauge-invariant operator of the form $(\bar L_\alpha i \sigma_2 H^*)(H^\dagger i \sigma_2 L_\beta^c)\phi$, where $L, H$ are the SM lepton and Higgs doublets.
The Higgs VEV breaks $SU(2)_L$ and projects the neutrino field out of $L$, leading to the neutrinophilic interaction.

In this work, we explore one-loop radiative corrections arising from the above self-interaction and the corresponding predictions relevant for precision electroweak measurements. With low-energy experimental input on the fine-structure and Fermi constants, the per-mille level measurement of $Z$ pole observables made by $e^+e^-$ colliders (such as LEP-II and SLC) has served as a powerful tool for testing the electroweak gauge symmetry~\cite{Hollik:1988ii} and constraining various new physics candidates~\cite{Cheng:2003ju, Ramsey-Musolf:2006evg,Haller:2018nnx,Harigaya:2023uhg,Bertuzzo:2025ejw}. We apply it to derive an important constraint on the neutrino self-interaction parameter space.

\begin{figure}[!t]
    \centering
    \includegraphics[width=0.48\linewidth]{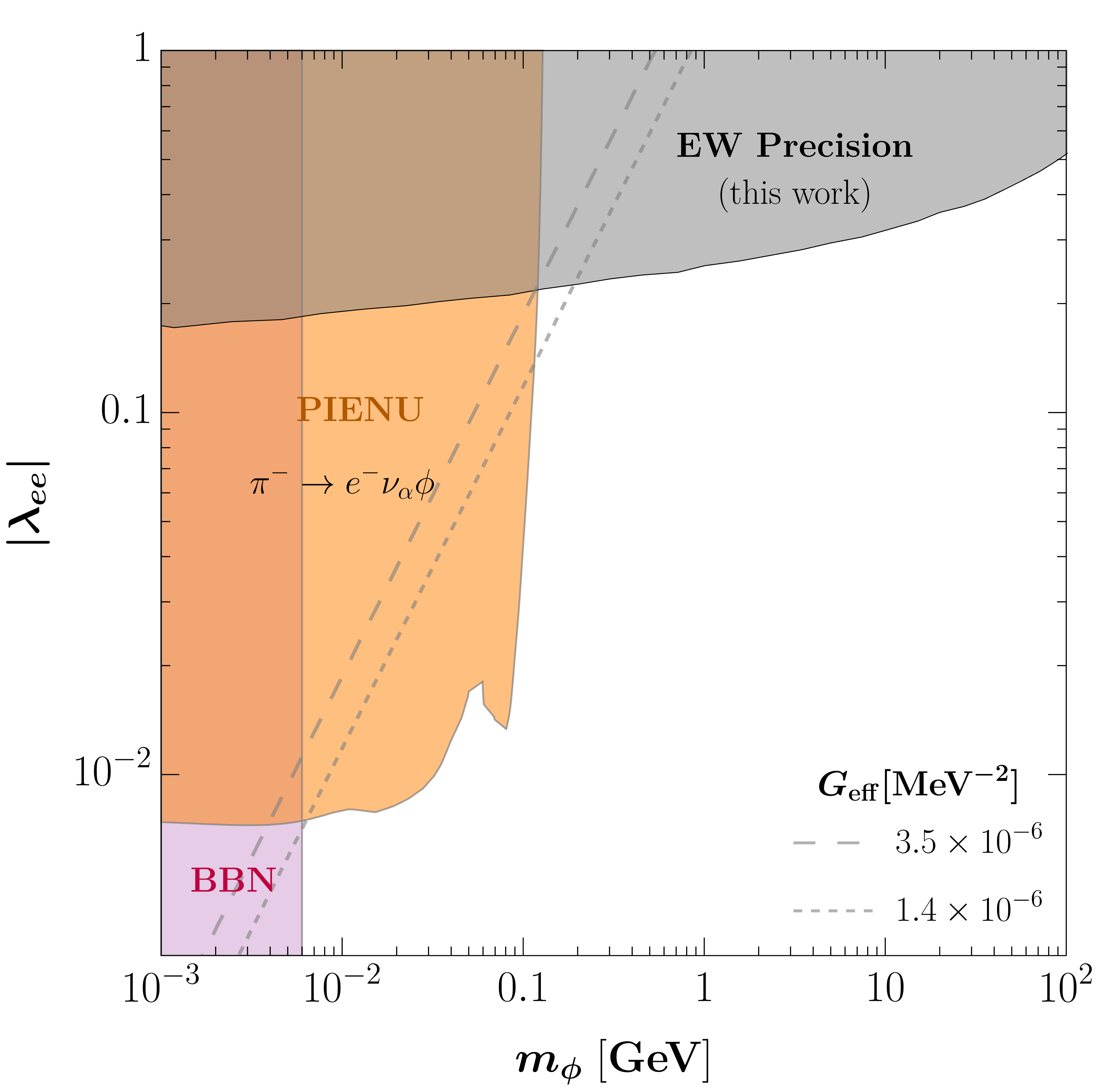}
    \includegraphics[width=0.48\linewidth]{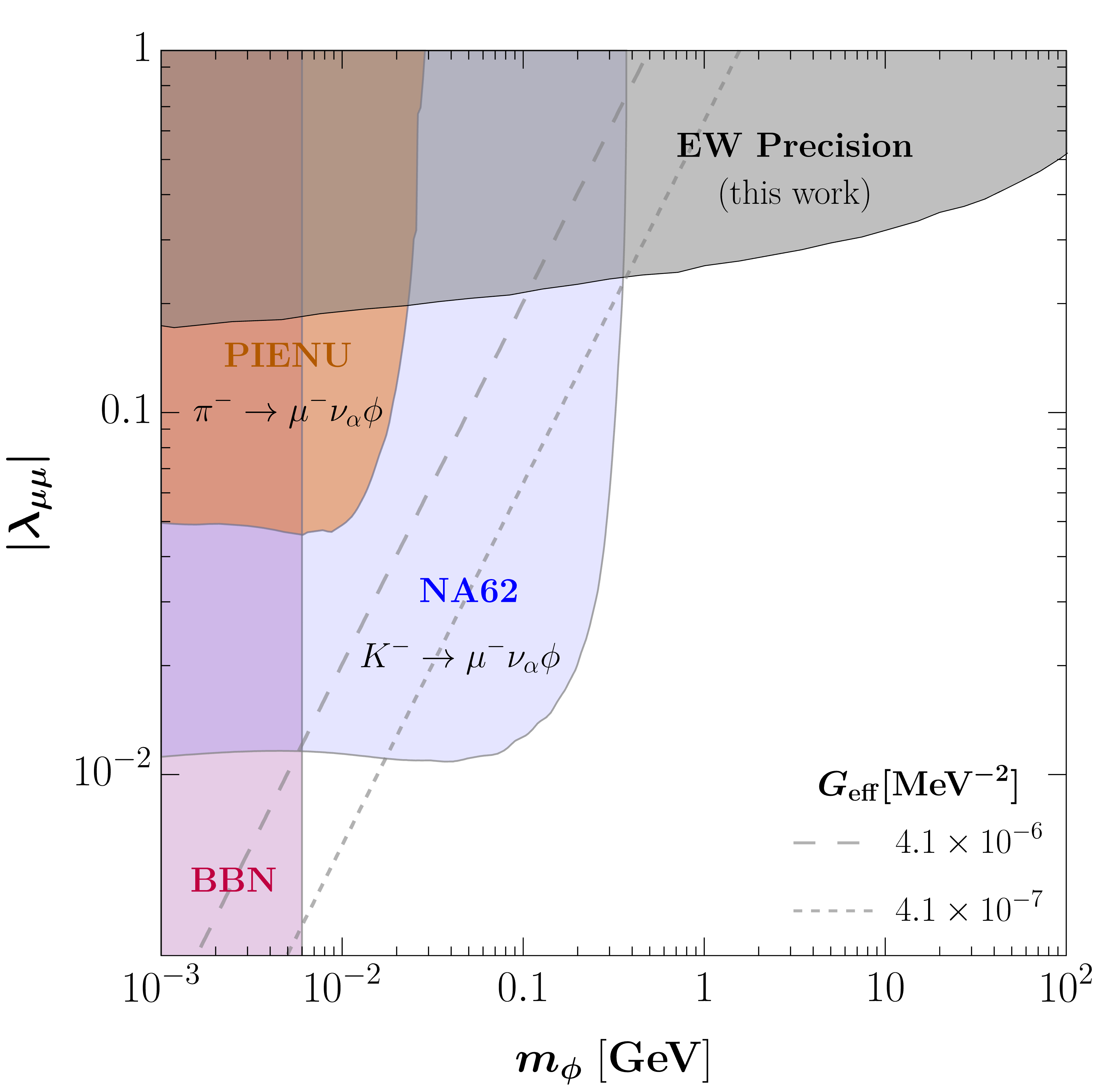}
    \includegraphics[width=0.48\linewidth]{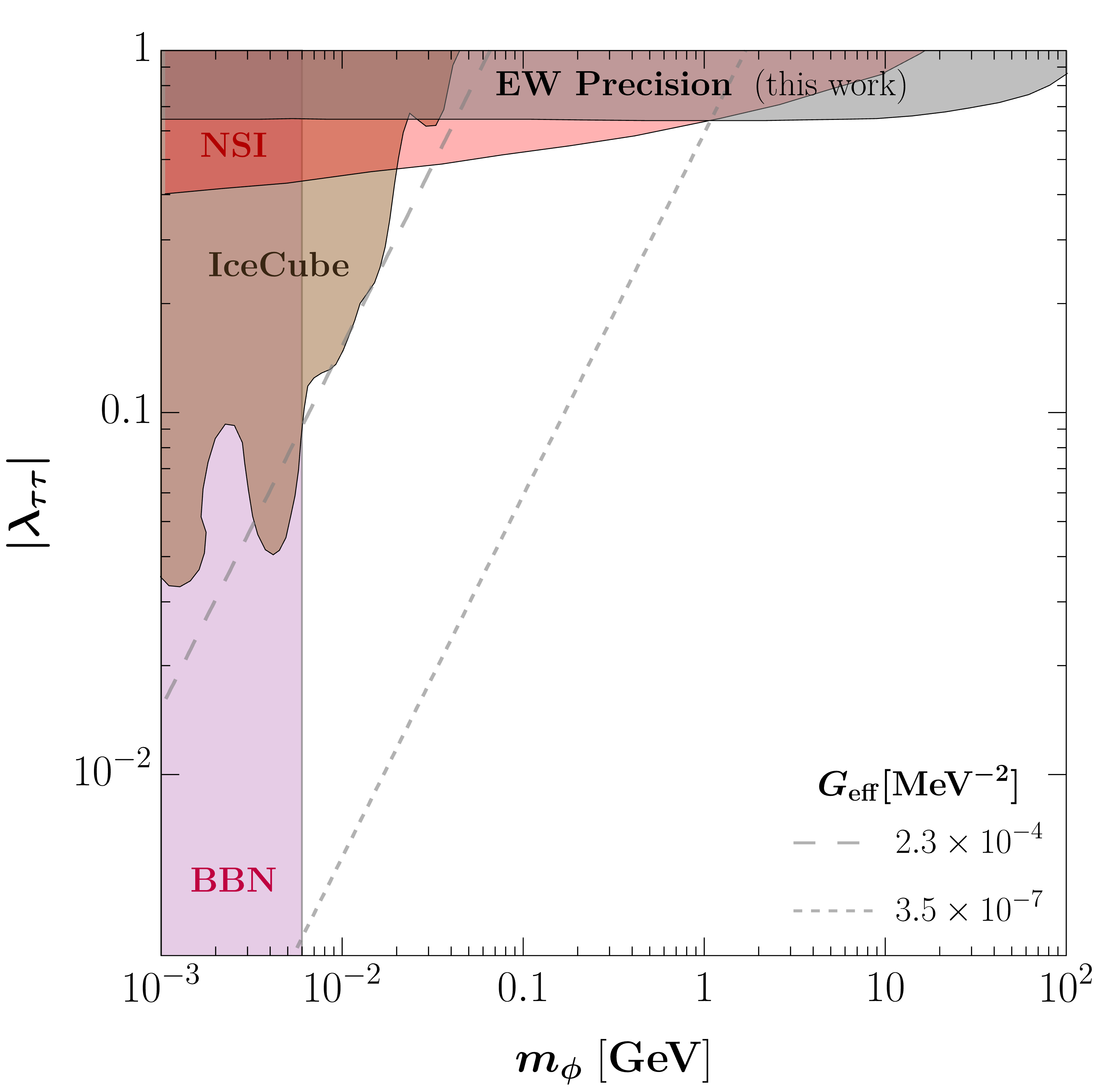}
    \includegraphics[width=0.48\linewidth]{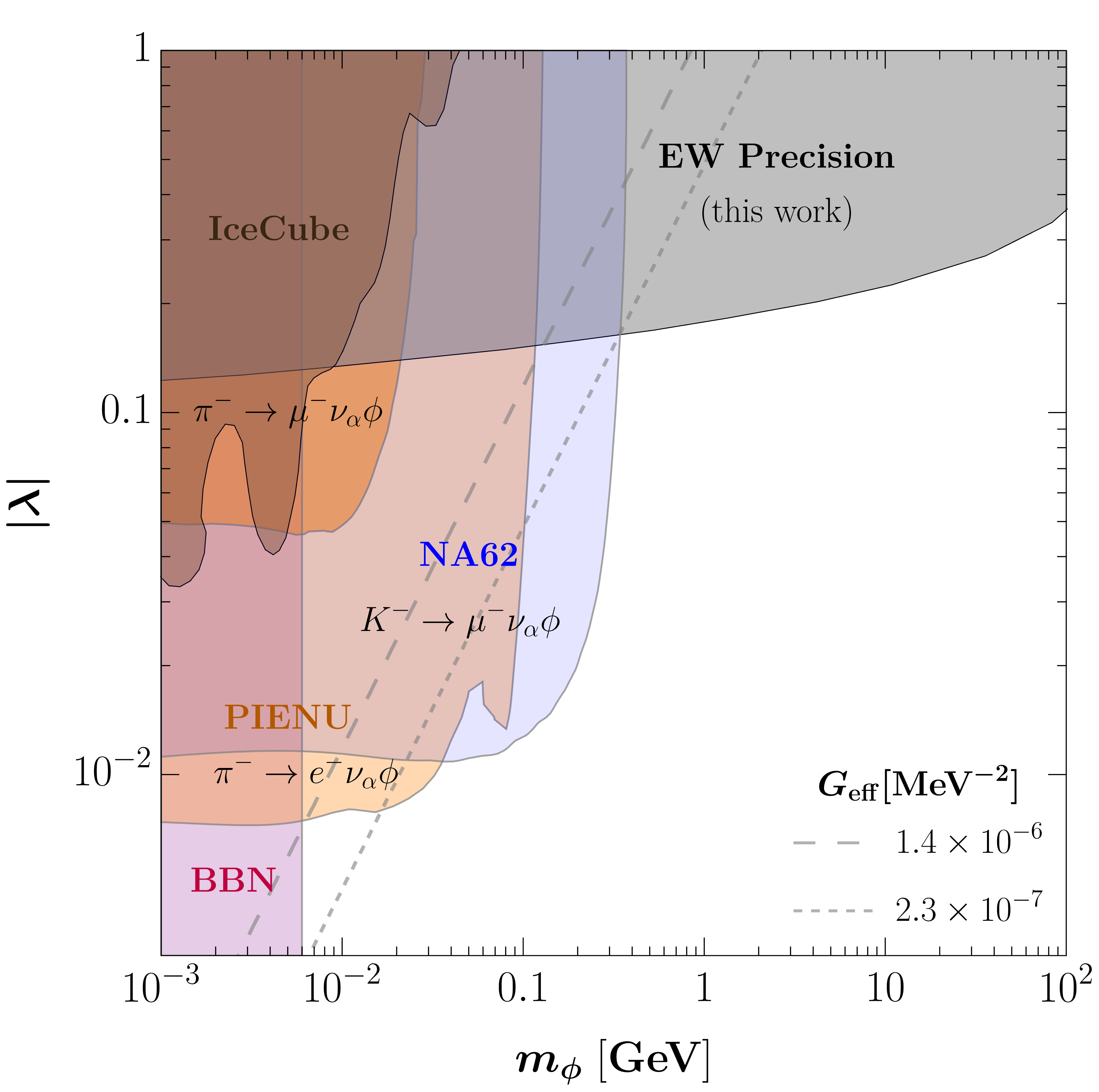}
    \caption{Constraints on neutrinophilic scalars mediating neutrino self-interaction for the flavor-specific and flavor-universal cases. The gray shaded regions excluded at the $2\sigma$ level from electroweak precision constraints are the main results of this work. The orange (blue) regions are ruled out from $\pi^-$ ($K^-$) decay constraints set by PIENU~\cite{PIENU:2021clt} (NA62~\cite{NA62:2021bji}).  The magenta shaded region represents the BBN bound: $m_\phi \geq 6$ MeV. In the $\nu_\tau$-specific case, the brown and red shaded regions correspond to the results derived from IceCube measurements~\cite{Esteban:2021tub}, and constraints from non-standard neutrino interactions~\cite{Foroughi-Abari:2025mhj}, respectively (see section~\ref{sec:lowenergy} for details). The gray dashed and dotted lines are the contours of constant $G_{\rm eff}\equiv\lambda^2/m_\phi^2$.}
    \label{fig:res}
\end{figure}

To compute the radiative corrections to electroweak gauge couplings, one must go beyond the simplified model Eq.~\eqref{eq:lagphi} by embedding it into a renormalizable and SM gauge invariant ultraviolet (UV) complete model. The cancellation of UV divergences is guaranteed by the Ward identity. The finite contribution can be divided into two parts, from loop diagrams involving $\phi$, and those involving only heavy particles in the UV completion. While both parts have to vanish if the heavy particle mass scale $M$ goes to infinity as dictated by the decoupling theorem, it was first suggested in~\cite{Zhang:2024meg} that the light $\phi$ contribution can be relatively more important due to a logarithmic enhancement factor, $\ln (M^2/m_\phi^2)$. The presence of such a log factor is independent of details of the UV completion.
We discuss its impact on electroweak physics in section~\ref{sec:ew}. 
In particular, in cases where $\phi$ couples to $\nu_e$ or $\nu_\mu$, the radiative correction modifies the Fermi constant extracted from muon decay and in turn affects $Z$-pole observations through the $\Delta r$ parameter. This allows us to set the leading constraint for $m_\phi$ above $\sim 300\,$MeV.
In section~\ref{sec:globalfit}, we perform a global fit to electroweak observables by altering the public {\tt Gfitter} code~\cite{Flacher:2008zq,Baak:2011ze} to constrain the $\lambda$ versus $m_\phi$ parameter space which is nearly model-independent in the light $\phi$ limit.
We review some existing limits from charged meson decay and neutrino experiments in section~\ref{sec:lowenergy}.
Together with the precision electroweak constraints, they allow us to chart the state-of-art probes of the self-interacting neutrino (SI$\nu$) scenario in Fig.~\ref{fig:res}.
In section~\ref{sec:UVcompletion}, we calculate the heavy particle loop contribution to electroweak observables in a concrete UV completion that includes $SU(2)_L$ triplet scalar and discuss conditions for it to be subdominant to the light $\phi$. We also point out that the electroweak oblique parameters can be used to further restrict the mass splitting among the triplet components, which is strongly correlated with the mass of light $\phi$.
We discuss the implication of our findings 
%for self-interacting neutrino cosmology (maximal allowed values of $G_{\rm eff}$) and sterile neutrino dark matter relic target in section~\ref{sec:implication} 
and conclude in section~\ref{sec:sum}.

%%%%%%%%%%%%%%%%%%%%%%%%%%%%%%%%%%%%%%%%%%%%%
\section{Impact of Neutrino Self-interaction on Electroweak Observables}
\label{sec:ew}
%%%%%%%%%%%%%%%%%%%%%%%%%%%%%%%%%%%%%%%%%%%%%

The $\phi$ mediated neutrino self-interaction can affect the neutral and charged current interactions of neutrinos through radiative corrections. The electroweak Lagrangian reads
\begin{align}\label{eq:EWL}
    \mathcal{L} = - (g_Z)_{\alpha\beta} \bar{\nu}_\alpha \gamma^\mu \mathbb{P}_L \nu_\beta Z_\mu - \left[ (g_W)_{\alpha\beta} \bar{\ell}_\alpha \gamma^\mu \mathbb{P}_L \nu_\beta W_\mu^- +\hc \rule{0mm}{4mm}\right]\,,
\end{align}
where $\mathbb{P}_{L,R} = (1\mp\gamma_5)/2$. At tree level, the couplings are flavor diagonal ($\alpha=\beta$), $g_Z^0 = g/(2 \cos{\theta_W})$, $g_W^0 = g/\sqrt{2}$, where $\theta_W$ denotes the weak mixing angle and $g$ is the $SU(2)_L$ gauge coupling. 

\begin{figure}[h]
    \centering
    \begin{tikzpicture}
\begin{feynman}
\vertex [label=above:\(Z\)] (a) at (0,0);
\vertex (b) at (0.7,0);
\vertex (c) at (1.57,0.5);
\vertex (d) at (1.57,-0.5);
\vertex [label=right:\(\nu_\alpha\)] (e) at (2.27,0.5);
\vertex [label=right:\(\nu_\beta\)] (f) at (2.27,-0.5);
\feynmandiagram [inline=(a.base)] {
(a)  -- [photon, momentum'=\(q\)] (b),
(b) -- [fermion, edge label=\(\nu^c_\gamma\)] (c),
(d) -- [charged scalar, edge label'=\(\phi\)] (c),
(d) -- [fermion, edge label=\(\nu^c_\gamma\)] (b),
(c) -- [fermion] (e),
(d) -- [anti fermion] (f),
};
\end{feynman}
\end{tikzpicture}
\hspace{3em}
\feynmandiagram [horizontal=a to d] {
  i1 [particle=\(\nu_\beta\)] -- [fermion, momentum'=\(p\)] a -- [fermion, edge label'=\(\nu^c_\gamma\)] b -- [fermion] i2 [particle=\(\nu_\alpha\)],
  a -- [charged scalar, half left, looseness=1.5, edge label=\(\phi\)] b,
}; 
    \caption{Feynman diagrams through which a light neutrinophilic $\phi$ affects the neutrino neutral and charged current interactions at one-loop level.}
    \label{fig:EWcurrent}
\end{figure}
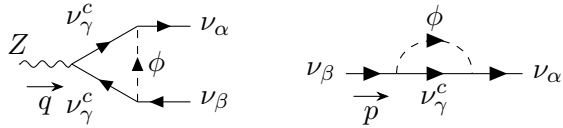

The presence of a neutrinophilic scalar $\phi$ contributes to the radiative corrections through the vertex and self-energy diagrams in Fig.~\ref{fig:EWcurrent}. 
Because Eq.~\eqref{eq:lagphi} is not gauge invariant under $SU(2)_L\times U(1)_Y$, the light $\phi$ contribution alone has UV divergences. 
This problem can be solved by resorting to a gauge invariant UV complete model that contains new heavy particles (with mass scale $M$ above the electroweak scale).
Inclusion of the new particle loop diagrams renders the result UV finite.
Regardless of the form of UV completion, the light $\phi$ contributions to neutrino electroweak couplings take the generic form~\cite{Zhang:2024meg} 
\begin{equation}\label{eq:dgzdgw}
\begin{split}
(\delta g_Z)_{\alpha\beta} &= - \frac{g_Z^0\lambda_{\alpha\gamma}\lambda^*_{\beta\gamma}}{32\pi^2}\Bigg[ \ln\frac{M^2}{m_\phi^2} + c_Z + f_1 (q^2, m_\phi, M) + \mathcal{O}\left(\frac{1}{M^2}\right) \Bigg] \ , \\
(\delta g_W)_{\alpha\beta} &= - \frac{g_W^0 \lambda_{\alpha\gamma}\lambda^*_{\beta\gamma}}{64\pi^2} \left[ \ln\frac{M^2}{m_\phi^2} +c_W + \mathcal{O}\left(\frac{1}{M^2}\right) \right] \ ,
\end{split}
\end{equation}
where $q^\mu$ is the momentum carried by the $Z$ boson. The $\mathcal{O}(1/M^2)$ terms are negligible for $|q^2|, m_\phi^2 \ll M^2$. 
The dummy flavor index $\gamma$ is summed over.

Both $\delta g_Z$ and $\delta g_W$ contain a logarithmic factor which can be large if $m_\phi$ lies well below the electroweak scale. The constants $c_Z, c_W$ are order one numbers whose exact values depend on the details of the UV completion. In the scalar triplet model to be discussed below, it has been found that $c_Z=-11/2$ and $c_W=-2$~\cite{Foroughi-Abari:2025mhj}. For light $\phi$ they are subdominant to the log factor. 
The finite loop function $f_1$ reads
\begin{equation}\label{eq:f1}
f_1 (q^2, m_\phi, M) \equiv 2{\rm Re}\int_0^1 dx \int_0^{1-x} dy \Bigg[\ln \left(\frac{M^2}{(1-x-y)m_\phi^2-xy q^2}\right)  + \frac{x y q^2}{(1-x-y)m_\phi^2-xy q^2} \Bigg] \ .
\end{equation}
The $q^2$ dependence in the neutral-current coupling $\delta g_Z$ and the interplay with $m_\phi^2$ has been explored in~\cite{Zhang:2024meg, Foroughi-Abari:2025mhj}.
In this work, we point out the importance of the correction $\delta g_W$ in the precision electroweak constraints. 
We will consider flavor diagonal $\lambda$ couplings  ($\alpha=\beta$) throughout this work. 
%This choice suppresses the dangerous lepton flavor violating processes that could occur at tree level in the UV complete model.
As a result, the gauge couplings $g_{W,Z}$ stay flavor diagonal.

Besides Eq.~\eqref{eq:dgzdgw}, there could be additional contributions to $\delta g_{Z,W}$ from heavy particle loops that do not involve $\phi$. We will address their importance in section~\ref{sec:UVcompletion} in the context of a concrete UV complete model.

\subsection{The Fermi Constant}\label{sec:GF}

The Fermi constant is defined as $G_F \equiv \sqrt{2} g^2/(8 m_W^2)$ and is most precisely determined by measuring the muon lifetime $\tau_\mu$.
For nonzero $\lambda_{ee}$ and/or $\lambda_{\mu\mu}$, the correction to $g_W$ affects the charged-current process. It is useful to define the new physics contribution to the $\Delta r$ parameter
 \begin{align}\label{eq:defDeltar}
    \Delta r^{\rm SI\nu} \equiv \frac{(\delta g_W)_{ee} + (\delta g_W)_{\mu\mu}}{g_W^0} = - \frac{|\lambda_{ee}|^2 + |\lambda_{\mu\mu}|^2}{64\pi^2} \left( \ln \frac{M^2}{m_\phi^2} + c_W \right) \ . 
\end{align}
The corresponding muon lifetime reads
 \begin{align}\label{eq:muonlife}
    \frac{1}{\tau_\mu} \simeq \left( 1 + \Delta r^{\rm SI\nu} \right)^2 \left( \frac{1}{\tau_\mu} \right)_{\rm SM}  \ .
\end{align}
This expression is precise for $m_\phi>m_\mu$. For $m_\phi \ll m_\mu$, a four-body muon decay channel opens up where $\phi$ can be radiated from a final state neutrino. In the $m_\phi\to0$ limit, it needs to be added to the three-body decay rate to render the total width finite.~\footnote{The $m_\phi\to0$ limit and infrared safety is only an academic question given the $\Delta N_{\rm eff}$ constraint from big-bang nucleosynthesis which requires $m_\phi$ to be above MeV scale.} 
Our precision electroweak analysis does not include the four-body decay contribution because for the mass range $m_\phi<m_\mu$, other constraints (from the exotic decays of charged meson to $\phi$) are much stronger, as shown in Fig.~\ref{fig:res}.

For nonzero $\Delta r^{\rm SI\nu}$, the extracted $G_F$ value from muon lifetime depends on the strength of neutrino self-interaction
\begin{align}\label{eq:gfsm2}
    G_F = \frac{G_\mu}{1 + \Delta r^{\rm SI\nu}} \simeq (1 - \Delta r^{\rm SI\nu}) G_\mu \ ,
\end{align}
where the second step is valid because $|\Delta r^{\rm SI\nu}|\ll1$. 
Throughout this work, we define
\begin{align}\label{eq:Gmu}
G_\mu \equiv 1.16637\times10^{-5}\,{\rm GeV}^{-2} \ .
\end{align}
This value is fixed by the muon lifetime measurement, which contains both the SM Fermi constant and the new physics correction $\Delta r^{\rm SI\nu}$.

In the SM limit, $\Delta r^{\rm SI\nu}=0$, thus $G_F=G_\mu$.
It is useful to note from Eq.~\eqref{eq:defDeltar} that $\Delta r^{\rm SI\nu}$ is negative for very light $\phi$.
Therefore, the presence of neutrino self-interaction increases the value of $G_F$ extracted from muon lifetime measurement.

It is also useful to note that if both $\lambda_{ee}$ and $\lambda_{\mu\mu}$ are nonzero, the $\phi$ exchange can induce an exotic muon decay through a box diagram as shown below. Because $\phi$ carries lepton number, the decay final state is $e^- \nu_e \bar\nu_\mu$. As a result, this box diagram does not interfere with the tree-level one. The corresponding correction to $G_F$ occurs at $\mathcal{O}(\lambda^4)$ and is subdominant to $\Delta r^{\rm SI\nu}$ which is $\mathcal{O}(\lambda^2)$. We do not take it into account in our analysis.
\begin{figure}[h]
    \centering
    \begin{tikzpicture}
       \begin{feynman}
\vertex [label=left:\(\mu^-\)] (a) at (0,0);
\vertex (b) at (1.0,0);
\vertex (c) at (1.75, 0.75); 
\vertex (d) at (1.75, -0.75);
\vertex (g) at (2.5, 0.0);
\vertex [label=right:\(e^-\)] (e) at (2.5,-1.5);
\vertex [label=right:\(\bar{\nu}_\mu\)] (f) at (2.5,1.5);
\vertex [label=right:\(\nu_e\)] (h) at (3.35,0.85);
\feynmandiagram [inline=(a.base)] {
(a)  -- [fermion] (b),
(b) -- [fermion, edge label=\(\nu_\mu\)] (c),
(b) -- [photon, edge label'=\(W^-\)] (d),
(c) -- [charged scalar, edge label = $\phi$] (g),
(d)--[anti fermion, edge label' = \(\bar{\nu}_e\)] (g),
(d)--[fermion] (e),
(c) --[anti fermion] (f),
(g) --[fermion] (h),
};
\end{feynman} 
    \end{tikzpicture}
\end{figure}

The $G_F$ measurement using muon lifetime alone cannot constrain new physics.
However, there are other ways to precisely determine $G_F$ from the measurement of other electroweak observables. In the SM, the agreement is at the per mille level, which demonstrates the success of the $SU(2)\times U(1)$ gauge theory and tightly restricts the room for new physics~\cite{Marciano:1999ih}. Alternatively, one can use the $G_F$ extracted from muon lifetime as an input to construct other electroweak observables.
The information of neutrino self-interaction will be encoded in the prediction of those observables and impact the global fit to the experimental data.

\subsection{The Weak Mixing Angle}\label{sec:WMA}

In the SM, the weak mixing angle is related to the fine-structure constant, the Fermi constant, and the physical mass of $Z$ boson as~\cite{Marciano:1999ih}
\begin{align}\label{eq:WMA}
    \sin^2\theta_W^{\rm SM} \cos^2\theta_W^{\rm SM} = \frac{\pi \alpha}{\sqrt{2} G_\mu m_Z^2 (1-\Delta r^{\rm SM})} \ ,
\end{align}
where $\Delta r^{\rm SM}$ accounts for radiative corrections to this relationship~\cite{Sirlin:1980nh}. 
For clarity, both $\sin^2\theta_W$ and $\Delta r$ are defined using the on-shell renormalization scheme, which is consistent with the convention used by the {\tt Gfitter} package.
In the presence of neutrino self-interaction, we have to replace $G_\mu$ by $G_F$ given in Eq.~\eqref{eq:gfsm2} and the new weak mixing angle is related to its SM counterpart as
\begin{align}
    \sin^2\theta_W \cos^2\theta_W = \frac{\sin^2\theta_W^{\rm SM} \cos^2\theta_W^{\rm SM} }{1 - \Delta r^{\rm SI\nu}} \simeq \frac{\pi \alpha}{\sqrt{2} G_\mu m_Z^2 (1-\Delta r^{\rm SM} - \Delta r^{\rm SI\nu})}  \ .
\end{align}
With $|\Delta r^{\rm SI\nu}|\ll1$, we can further derive
\begin{align}\label{eq:WMAdr}
    \sin\theta_W  \simeq \sin\theta_W^{\rm SM} \left[ 1 + \frac{1 + \sec(2\theta_W^{\rm SM})}{4} \Delta r^{\rm SI\nu} \right] \ .
\end{align}

Eqs.~\eqref{eq:gfsm2} and \eqref{eq:WMAdr} represent the contribution of neutrino self-interaction to the electroweak sector through the $\Delta r$ parameter.
If this were the only effect, one could already apply the $2\sigma$ limit 
\begin{align}\label{eq:deltar}
|\Delta r^{\rm SI\nu }| < 0.0012\,\, , 
\end{align}
given in~\cite{Erler:2004cx} to set an upper bound on $\lambda$ as a function of $m_\phi$. However, as we will see below, neutrino self-interaction also impacts electroweak observables in other ways, through the radiative correction to $\delta g_Z$, as well as directly contributing to new decay channels with light $\phi$ in the final state.
A more careful analysis is required and presented below.

\subsection{$Z$-Pole Observables}\label{sec:Zpole}

The $Z$-boson pole observables have been measured to very high precision, primarily by the LEP and SLC experiments~\cite{ALEPH:2005ab,SLD:2000leq}. These include the mass ($m_Z$), the total decay width of $Z$-boson ($\Gamma_Z$), partial decay widths and their ratios ($R_i$), the asymmetry parameters ($A_i$), the total hadronic cross-section ($\sigma^0_{\rm had}$), and the effective weak mixing angle ($\sin^2\theta_{\rm eff}^{\ell}$). 
Their measured values are listed in the second column of Table~\ref{tab:fit}.
Here we discuss how they are affected by neutrino self-interaction.

\begin{itemize}
\item The $Z$ boson mass $m_Z = 91.188~\GeV$ is used as an input in our analysis.

\item Due to parity violation in the EW sector, the initial and final state fermion polarizations in neutral current interactions such as $e^+ e^- \rightarrow f\bar{f}$ have an observable asymmetry that is given by the asymmetry parameters
\begin{align}
    A_f &= \frac{2 g_{V,f}g_{A,f}}{g_{V,f}^2 + g_{A,f}^2}\,,
\end{align}
where $f$ denotes SM fermions, $f= e,\mu,\tau,b,c,s$, and
\begin{align}\label{eq:gz}
g_{V,f} = \sqrt{\rho_Z^f} \left( I_3^f - 2 Q^f \kappa_Z^f \sin^2\theta_W \right)\,, \quad g_{A,f} = \sqrt{\rho_Z^f} I_3^f \,,
\end{align}
are the vector and axial couplings of the $Z$-boson to fermions. The electroweak form factors $\rho_Z^f\,,\kappa_Z^f$, which are equal to unity at the tree level, encode the radiative corrections evaluated on the $Z$-pole~\cite{Marciano:1980pb}. 
In the simplified neutrino self-interaction model Eq.~\eqref{eq:lagphi}, the new corrections to $\rho_Z^f$ and $\kappa_Z^f$ occur beyond one-loop level due to the neutrinophilic nature of $\phi$, unless $f=\nu$.
%Note that these asymmetries correspond to a pure $Z$ exchange and only the real part of these couplings are relevant. 
The leptonic asymmetries are combined into a single observable $A_l$ assuming lepton flavour universality, leaving us with 4 observables: $A_l,A_b,A_c$ and $A_s$.
These asymmetry parameters are affected by $\Delta r^{\rm SI\nu}$ through the weak mixing angle Eq.~\eqref{eq:WMAdr}.

\item The above asymmetry parameters are used to construct the forward-backward asymmetry as
\begin{align}
A_{\rm FB}^{f}  = \frac{\sigma_F^f - \sigma_B^f}{\sigma_F^f + \sigma_B^f} =& \frac{3}{4} A_e A_f\,,
\end{align}
where $\sigma_{F,B}^f$ is the fermionic production cross-section in the forward/backward hemisphere relative to the electron beam direction. 
Similarly, $A_{\rm FB}^{f}$ is also affected by neutrino self-interaction through the weak mixing angle. 

\item The effective weak mixing angle ($\sin^2\theta_{\rm eff}^{\ell}$) on the $Z$-pole is related to the $\sin^2\theta_W$ in Eq.~\eqref{eq:WMAdr} as~\cite{Gambino:1993dd}
\begin{align}
\sin^2\theta_{\rm eff}^{\ell} = \kappa_Z^\ell \sin^2\theta_W \ .
\end{align}
It has been measured using charge flow ($Q_{\rm FB}$) at LEP~\cite{ALEPH:2005ab} and recently at hadron colliders (HC) including the Tevatron and LHC~\cite{CDF:2018cnj, ATLAS:2018gqq, CMS:2024ony, LHCb:2015jyu}. 

\item The $Z$ boson total decay width is given by
\begin{align}\label{eq:GammaZ}
\Gamma_Z = \Gamma_Z^{\rm SM}\left(\Delta r^{\rm SI\nu}\right) + \delta \Gamma_{Z\to\nu\bar\nu} + \left(\Gamma_{Z\to\nu\nu\phi^*} + \Gamma_{Z\to\bar\nu\bar\nu\phi}\right) \ ,
\end{align}
where the SM decay width is given by $\Gamma_Z^{\rm SM} = \Gamma_e + \Gamma_\mu + \Gamma_\tau + 3 \Gamma_\nu + \Gamma_{\rm had}$ and $\Gamma_{\rm had} = \Gamma_u + \Gamma_d+\Gamma_s+\Gamma_c+\Gamma_b$.
The partial decay widths are written in terms of $G_F$ and $\sin\theta_W$ (see Eq.~(21) in Ref.~\cite{Flacher:2008zq}). In the presence of neutrino self-interaction, both $G_F$ and $\sin\theta_W$ are affected by $\Delta r^{\rm SI\nu}$ as given by Eqs.~\eqref{eq:gfsm2} and \eqref{eq:WMAdr}. 

\item The neutrino final state requires special care in the presence of self-interaction.
In Eq.~\eqref{eq:GammaZ}, the second term on the right-hand side, $\delta \Gamma_{Z\to\nu\bar\nu}$ accounts for the shift to $Z\to\nu\bar\nu$ decay width (summing over all flavors) due to the radiative correction to neutrino neutral current coupling, $\delta g_Z$ given by Eq.~\eqref{eq:dgzdgw}. 
Essentially, $\delta g_Z$ shifts the electroweak form factor $\rho_Z^\nu$ as
\begin{align}
\left( \sqrt{\rho_Z^{\nu_\alpha}} - 1 \right)_{\rm SI\nu} = \frac{{\rm Re} (\delta g_Z)_{\alpha\alpha}}{g_Z^0} \ .
\end{align}
This partial width shift has been calculated in~\cite{Foroughi-Abari:2025mhj} (see also~\cite{Foroughi-Abari:2025upe}) and
\begin{equation} \label{eq:shiftZ2body}
\begin{split}
    \delta \Gamma_{Z\to\nu\bar\nu} &\simeq \Gamma_{Z\to\nu_\alpha \bar\nu_\alpha}^\text{SM, tree} \cdot \frac{2{\rm Re}(\delta g_Z)_{\alpha\alpha}}{g_Z^0} \\
    &= - \frac{\sqrt{2} G_\mu m_Z^3 |\lambda_{\alpha\alpha}|^2}{384\pi^3} \Bigg[ \ln\frac{M^2}{m_\phi^2} + {\rm Re}\, f_1 \left(q^2=m_Z^2, m_\phi, M \right) + c_Z  \Bigg] \ ,
\end{split}
\end{equation}
where $\Gamma_{Z\to\nu_\alpha \bar\nu_\alpha}^\text{SM, tree} = G_\mu m_Z^3/(12\sqrt{2} \pi)$ and the flavor index $\alpha=e,\mu,\tau$ is summed over.
Here, we do not have to distinguish between $G_F$ and $G_\mu$ because the contribution is already at $\mathcal{O}(\lambda^2)$ order.
The last bracket of Eq.~\eqref{eq:GammaZ} is a new three-body decay channel of the $Z$ boson with an on-shell $\phi$ (or $\phi^*$) in the final state. Because $\phi$ eventually decays into neutrinos, the final state is also invisible. The three-body decay rate is~\cite{Foroughi-Abari:2025mhj}
\begin{align}\label{eq:shiftZ3body}
 &\Gamma_{Z\to\nu\nu \phi^*} + \Gamma_{Z\to\bar\nu\bar\nu\phi} = \frac{\sqrt{2} G_\mu m_Z^3 |\lambda_{\alpha\alpha}|^2}{192\pi^3} \left\lbrace \left(\ln{\frac{m_Z^2}{m_\phi^2}} - \frac{11}{3}\right) + \frac{8m_\phi^2}{m_Z^2}\left(\ln{\frac{m_Z^2}{m_\phi^2}} - \frac{5}{8}\right) \right. \nonumber\\
& \left.+ \frac{2m_\phi^4}{m_Z^4} \left[\ln{\frac{m_Z^2}{m_\phi^2}} \left(
\ln{\frac{m_Z m_\phi}{m_Z^2+m_\phi^2}} + \frac{3}{2}\right) + \frac{9}{2} 
+{\rm Li}_2\left( \frac{m_Z^2}{m_Z^2+m_\phi^2}\right)
-{\rm Li}_2\left( \frac{m_\phi^2}{m_Z^2+m_\phi^2}
\right)
\right] - \frac{m_\phi^6}{3m_Z^6}\right\rbrace \ .
\end{align}
Again, the flavor index $\alpha$ is summed over. 
In the $m_\phi \ll m_Z$ limit, 
\begin{align} 
    \delta \Gamma_{Z\to\nu\bar\nu}  + \left(\Gamma_{Z\to\nu\nu \phi^*} + \Gamma_{Z\to\bar\nu\bar\nu\phi}\right) \simeq - \frac{G_\mu m_Z^3 |\lambda_{\alpha\alpha}|^2}{96\sqrt{2}\pi^3} \left( \ln\frac{M^2}{m_Z^2} + \frac{17+3 c_Z}{6} \right) \ ,
\end{align}
which is independent of $m_\phi$.  If $\Delta r^{\rm SI\nu}\neq0$, the remaining $\ln (M^2/m_\phi^2)$ enhancement resides in the first term of Eq.~\eqref{eq:GammaZ} which makes the dominant correction to the total $Z$ width.

\item The remaining observables are the partial width ratios and total hadronic cross-section at $Z$ pole, defined as
\begin{align}
    R_\ell = \frac{\Gamma_{\rm had}}{\Gamma_\ell}\,,\quad R_q =  \frac{\Gamma_q}{\Gamma_{\rm had}}\,,\quad \sigma_{\rm had} = \frac{12\pi}{m_Z^2} \frac{\Gamma_e \Gamma_{\rm had}}{\Gamma_Z^2}\,,
\end{align}
where $q = b,c,s$. In the presence of neutrino self-interaction, all of these partial decay widths, and in turn $R_\ell$, $R_q$ and $\sigma_{\rm had}$, receive correction from $\Delta r^{\rm SI\nu}$ through $G_F$ and $\sin\theta_W$.

\end{itemize}

\subsection{Non-$Z$-pole Observables}\label{sec:nonZpole}

The two main $W$-boson observables are its mass and decay width determined from the combination of the experiments at Tevatron and LHC~\cite{CDF:2012gpf,D0:2012kms,ATLAS:2024erm}. 
Their values are given in the second column of Table~~\ref{tab:fit}.

The physical masses of $W,Z$ bosons are used to define the weak mixing angle 
\begin{equation}\label{eq:MW}
m_W \equiv m_Z \cos\theta_W \ ,
\end{equation}
$\theta_W$ is given by Eq.~\eqref{eq:WMAdr} and depends on neutrino self-interaction through $\Delta r^{\rm SI\nu}$, and with $m_Z$ as input, so does the $W$ mass.
Under the on-shell renormalization scheme, the above relation holds to all orders of perturbation theory~\cite{Hollik:1988ii}.

Similar to $Z$ boson, the total decay width of the $W$ boson can be written as
\begin{align}\label{eq:GammaW}
\Gamma_{W^-} = \Gamma_{W^-}^{\rm SM}\left(\Delta r^{\rm SI\nu}\right) + \delta \Gamma_{W^-\to\ell\bar\nu} + \Gamma_{W^-\to\ell\nu\phi^*} \ .
\end{align}
The SM $W$ decay width is written in terms of $G_F$ and $m_W$ (see Eq.~(20) of~\cite{Cho:2011rk}). In the presence of neutrino self-interaction, both are affected by $\Delta r^{\rm SI\nu}$ via Eqs.~\eqref{eq:gfsm2} and \eqref{eq:MW}.

The radiative correction to $g_W$ makes an additional contribution to the leptonic decay amplitude of the $W$ boson, denoted as $\delta \Gamma_{W^-\to\ell\bar\nu}$ in Eq.~\eqref{eq:GammaW}. We find
\begin{equation}\label{eq:shiftW2body}
\begin{split}
    \delta\Gamma_{W \rightarrow \ell\bar\nu} \simeq \Gamma_{W \rightarrow \ell\bar\nu}^\text{SM, tree} \cdot \frac{2(\delta g_W)_{\alpha\alpha}}{g_W^0} = - \frac{G_\mu m_W^3 |\lambda_{\alpha \alpha}|^2}{192\,\sqrt{2}\pi^3} \left(\ln{\frac{M^2}{m_\phi^2}}+c_W\right)\ ,
\end{split}
\end{equation}
where $\Gamma_{W \rightarrow \ell\bar\nu}^\text{SM, tree} = G_\mu m_W^3/(6\sqrt{2}\pi)$. 
A light $\phi$ can also be radiated from final state neutrino, leading to a three-body decay mode of the $W$ boson
\begin{align}\label{eq:shiftW3body}
    \Gamma_{W \rightarrow \ell\nu\phi^*} &= \frac{G_\mu m_W^3 |\lambda_{\alpha \alpha}|^2}{1152\,\sqrt{2}\pi^3} \left[-17 + 9\left(\frac{m_\phi^2}{m_W^2} + \frac{m_\phi^4}{m_W^4}\right)+ 6\left(1+3 \frac{m_\phi^2}{m_W^2}\right)\ln{\frac{m_W^2}{m_\phi^2}}-\frac{m_\phi^6}{m_W^6}\right]\,,
\end{align}
which is kinematically allowed for $m_\phi < m_W$, and we have neglected the charged lepton masses.
In both Eqs.~\eqref{eq:shiftW2body} and \eqref{eq:shiftW3body}, the flavor index $\alpha$ is summed over $e,\mu,\tau$. 
In the $m_\phi\ll m_W$ limit, 
\begin{align}
\delta\Gamma_{W \rightarrow \ell\nu} + \Gamma_{W \rightarrow \ell\nu\phi} \simeq - \frac{G_\mu m_W^3 |\lambda_{\alpha \alpha}|^2}{192\,\sqrt{2}\pi^3} \left( \ln\frac{M^2}{m_W^2} + \frac{17 + 2 c_W}{6} \right) \ .
\end{align}
The $\ln(M^2/m_\phi^2)$ enhancement factor is absent here, similar to the case of $Z$ decay.

%%%%%%%%%%%%%%%%%%%%%%%%%%%%%%%%%%%%%%
\section{Precision Electroweak Constraint on Simplified Model}
\label{sec:globalfit}
%%%%%%%%%%%%%%%%%%%%%%%%%%%%%%%%%%%%%%

The global fit of the predictions to the electroweak precision data is not only instrumental for validating the gauge structure of the SM but also provides powerful tests of beyond the SM physics.
Given the radiative corrections to both charged- and neutral-current interactions from a neutrinophilic scalar, it becomes even more pertinent to use a fitting program to determine the precision constraints, instead of just constraining the $Z$ decay width as was done in previous works~\cite{Brdar:2020nbj, Foroughi-Abari:2025mhj}. 

We adopt and modify the $\gfit$ package~\cite{Flacher:2008zq,Baak:2011ze} to include the effects from the self-interacting neutrino sector, including the simplified model that contains a neutrinophilic scalar, as well as a UV completion to be discussed in section~\ref{sec:UVcompletion}.
The package follows a frequentist approach to find the global minimum of square ($\chi^2_{\rm min}$), with the test statistic for a set of model parameters $y_i$ defined by
\begin{align}
    \chi^2 \left(y_i\right) \equiv - 2\ln\mathcal{L}(y_i)\,,
\end{align}
where $\mathcal{L}$ is the likelihood function.

\subsection{Global Fit for the SM}

We begin with a telegraphic review of the SM global fit. 
The following input parameters are allowed to float within their experimentally determined range
\begin{align}
    y_i =\{\alpha_s(m_Z^2),~\Delta \alpha_{\rm had}^{(5)}(m_Z^2),~m_Z,~ m_t,~ M_H\}\,,
\end{align}
where $\alpha_s(m_Z^2)$ is the strong coupling constant, $\Delta \alpha_{\rm had}^{(5)}$ is the hadronic contribution of the light five quarks to the electromagnetic coupling constant (evaluated at the $m_Z$ scale), $m_t$ is the top mass and $M_H$ is the SM Higgs boson mass. Along with these parameters, the other floating parameters are the quark masses: $\{m_c,m_b\}$. The lighter quark masses have been precisely determined by lattice QCD~\cite{BMW:2010ucx, FermilabLattice:2018est} and are kept fixed in the fit, along with $G_F=G_\mu$ measured using the muon lifetime. 

The $\gfit$ code incorporates theoretical computations up to full two-loop and beyond-two-loop corrections for $m_W,\sin^2\theta_{\rm eff}^\ell$, partial and total $Z$ decay widths~\cite{Awramik:2003rn,Awramik:2006uz,Freitas:2014hra,Cho:2011rk}. The missing higher order corrections are all encapsulated via theoretical uncertainty parameters~\cite{ParticleDataGroup:2024cfk,Flacher:2008zq}
\begin{align}
    &\dth m_W = 4~\MeV,\,\dth \sin^2\theta_{\rm eff}^{f} = 4.7 \times 10^{-5},\,\dth \Gamma_Z = 0.4~\MeV,\,\dth \sigma_{\rm had} = 6~{\rm pb}\,,\nn\\
    &\dth R_\ell = 0.6,\,, \dth R_b = 0.0001,\, \dth R_c = 0.00005 \ .
\end{align}

\begin{table}[!t]
\centering
\footnotesize
\renewcommand{\arraystretch}{1.1}
\begin{tabular}{c c | c c | c c}
\toprule
& & \multicolumn{2}{c|}{SM} & \multicolumn{2}{c}{SI$\nu$ (Universal)}\\
\hline
Parameter & Input value & Fit & Pull & Fit & {Pull} \\
\hline
$\alpha_s(m_Z^2)$ & $0.1177 \pm 0.0009$ & $0.118\pm 0.0009$ & $+0.37$ & $0.1180\pm 0.0008$ & $+0.33$ \\
$\Delta\alpha_{\rm{had}}^{(5)}(m_Z^2)^{(\Delta)}$ & $0.02758 \pm 0.00010$ & $0.02757\pm 0.00009$ & $-0.17$ & $0.02756\pm 0.0001$ & $-0.27$ \\
$m_Z$ [GeV] & $91.188 \pm 0.0020$ & $91.1885\pm 0.0020$ & $+0.25$ & $91.1885 \pm 0.0020$ & $+0.25$ \\
$m_t$ [GeV] & $172.56 \pm 0.61$ & $172.70\pm 0.58$ & $+0.20$ & $172.81 \pm 0.8$ & $+0.35$ \\
$M_H$ [GeV] & $125.1 \pm 0.1$ & $125.1\pm 0.1$ & $0.00$ & $125.1 \pm 0.00$ & $0.00$ \\
\hline
$m_W$ [GeV] & $80.369 \pm 0.013$ & $80.358\pm 0.006$ & $-0.86$ & $80.362\pm 0.007$ & $-0.55$ \\
$\Gamma_W$ [GeV] & $2.140 \pm 0.050$ & $2.090\pm 0.001$ & $-1.00$ &$2.091\pm 0.001$ & $-0.98$ \\
\hline
$\sin^2\theta_{\rm{eff}}^\ell(\rm{HC})$ & $0.23152 \pm 0.00023$ & $0.23152\pm 0.00006$ & $0.00$ & $0.23152\pm 0.00006$ & $+0.26$ \\
$\sin^2\theta_{\rm{eff}}^\ell(Q_{\rm{FB}})$ & $0.2324 \pm 0.0012$ & $0.23152\pm 0.00006$ & $-0.73$ & $0.23152\pm 0.00006$ & $-0.68$  \\
$A_\ell(\mathrm{LEP})$ & $0.1465 \pm 0.0033$ & $0.1470\pm 0.0004$ & $+0.15$ & $0.1470\pm 0.0004$ & $-0.02$ \\
$A_\ell(\mathrm{SLD})$ & $0.1513 \pm 0.0021$ & $0.1470\pm 0.0004$ & $-2.05$ & $0.1468\pm 0.0004$ & $-2.26$ \\
$\Gamma_Z$ [GeV] & $2.4955 \pm 0.0023$ & $2.4945\pm 0.0005$ & $-0.43$ & $2.4940\pm 0.0006$ & $-0.65$ \\
$\sigma_{\mathrm{had}}$ [nb] & $41.480 \pm 0.033$ & $41.489\pm 0.008$ & $+0.27$ & $41.489\pm 0.004$ & $+0.27$ \\
$R_\ell$ & $20.767 \pm 0.025$ & $20.733\pm 0.006$ & $-1.35$ & $20.733\pm 0.005$ & $-1.38$ \\
$A_{\mathrm{FB}}^\ell$ & $0.0171 \pm 0.0010$ & $0.01620\pm 0.0001$ & $-0.90$ & $0.01610\pm 0.0002$ & $-1.00$ \\
$R_b$ & $0.21629 \pm 0.00066$ & $0.21582\pm 0.00003$ & $-0.70$ & $0.21583\pm 0.00003$ & $-0.71$ \\
$R_c$ & $0.1721 \pm 0.0030$ & $0.17220\pm 0.00003$ & $+0.04$ & $0.17222\pm 0.00002$ & $+0.04$\\
$A_{\mathrm{FB}}^b$ & $0.0996 \pm 0.0016$ & $0.1030\pm 0.0003$ & $+2.13$ & $0.1027\pm 0.0005$ & $+1.94$\\
$A_{\mathrm{FB}}^c$ & $0.0707 \pm 0.0035$ & $0.0736\pm 0.0002$ & $+0.83$ & $0.0733\pm 0.0004$ & $+0.74$ \\
$A_b$ & $0.923 \pm 0.020$ & $0.93462\pm 0.00003$ & $+0.58$ & $0.93458\pm 0.00006$ & $+0.58$\\
$A_c$ & $0.670 \pm 0.027$ & $0.6678\pm 0.00019$ & $-0.08$ & $0.6677\pm 0.0003$ & $-0.09$ \\
\hline
$|\lambda|$ & $[0.001 - 1.0]$ & $-$ & $-$ & $0.064$ & $-$ \\
$m_\phi$ [GeV] & $[0.001 - 200.0]$& $-$ & $-$ & 0.056 & $-$\\
\midrule
& $G_F$ & \multicolumn{2}{c|}{$1.166379 \times 10^{-5}$} & \multicolumn{2}{c}{$1.166649 \times 10^{-5}$}\\
\hline
& $\chi_{\rm min}^2$ & \multicolumn{2}{c|}{13.99} & \multicolumn{2}{c}{13.60}\\
\bottomrule
\end{tabular}
\caption{Results of the SM and SI$\nu$ fit (flavour-universal case) to the electroweak precision observables, we fix $M = 1$ TeV. The first column indicates the parameters, with the blocks representing the input (floating in the fit) parameters, $\wb$-observables, $Z$-boson pole observables, and the new physics parameters (also floating in the fit within their specified ranges), respectively. The second column indicates the experimental average values taken from PDG~\cite{ParticleDataGroup:2024cfk}. The third (fifth) column correspond to the SM (SI$\nu$) fit values resulting in $\chi^2_{\rm min} = 13.99~(13.60)$. The best-fit value of $|\lambda|$ and $m_\phi$, and their input range are provided in the second and the fifth column, respectively. The fourth (sixth) column correspond to the pull values: $(O_{\rm fit}- O_{\rm meas})/\sigma_{\rm meas}$ in the SM (SI$\nu$) case. $^{(\Delta)}$ denotes rescaling due to $\alpha_s$ dependency.}
\label{tab:fit}
\end{table}

All the electroweak observables contributing to the global fit, along with their input values, are given in first and second columns of Table~\ref{tab:fit}.  The experimental uncertainties represent $1\sigma$ deviation. The SM has in total 19 parameters out of which 5 are floating in the fit, thus leaving 14 degrees of freedom.  
For observables with two or more experimental values, we take the weighted average, therefore $\seff(\rm{Avg}) = 0.23155 \pm 0.00023$ and $A_\ell (\rm{Avg}) = 0.1499 \pm 0.0018$.
Performing a single global fit for the SM gives $\chi^2_{\rm min}/{n_{\rm dof}} = 13.99/14$, corresponding to a $p$-value, $P=45\%$. For $\chi^2$ distributions, $P$ is defined as
\begin{align}
    P\equiv\mathcal{P}(\chi^2_{\rm min},n)= \int_{\chi^2_{\rm min}}^\infty  \frac{x^{n/2-1}e^{-x/2} dx}{2^{n/2}\Gamma(n/2)}\,,
\end{align}
and $\Gamma$ denotes the gamma function. 
This quantity is an indicator of the quality of the best fit point.
The fit values and pulls are shown in the third and fourth columns of Table~\ref{tab:fit}.
Overall, the SM predictions fit the data extremely well, except for $2\sigma$ level tensions in $A_\ell$ (SLD) and $A_{\rm FB}^b$. 

\subsection{Global Fit for the Self-interacting Neutrino Scenario}

Next, we spell out our global analysis for the self-interacting neutrino model with neutrinophilic scalars. For the simplified model Eq.~\eqref{eq:lagphi}, three more parameters are present in addition to those in the SM,
\begin{align}
\{\lambda,m_\phi, M\} \ .
\end{align}
We will hold $M$ fixed around TeV, and allow the other two parameters to vary freely in the following windows, $|\lambda| \in [0.01, 1]$ and $m_\phi \in [0.001, 100]\,\GeV$.
In order to directly compare with the results in the UV complete model to be discussed in the next section, we set $c_Z=-11/2$, $c_W=-2$ in Eq.~\eqref{eq:dgzdgw}.
We take into account the modification to $G_F$ due to $\Delta r^{\rm SI\nu}$ which further shifts $m_W, \sin^2\theta_W$ and the corresponding electroweak observables as explained in sections~\ref{sec:GF} -- \ref{sec:nonZpole}. We also include Eqs~\eqref{eq:shiftZ2body}, \eqref{eq:shiftZ3body} and \eqref{eq:shiftW2body}, \eqref{eq:shiftW3body}, which are direct contributions to $W$ and $Z$ partial decay rates into neutrino final states. Results of the global fit are presented as upper bound on $\lambda$ as a function of $m_\phi$, the gray excluded regions in Fig.~\ref{fig:res}. 
They serve as a main result of this work.

\subsubsection{Flavor-Universal Case}

\begin{figure}[!t]
    \centering
    \includegraphics[width=0.48\linewidth]{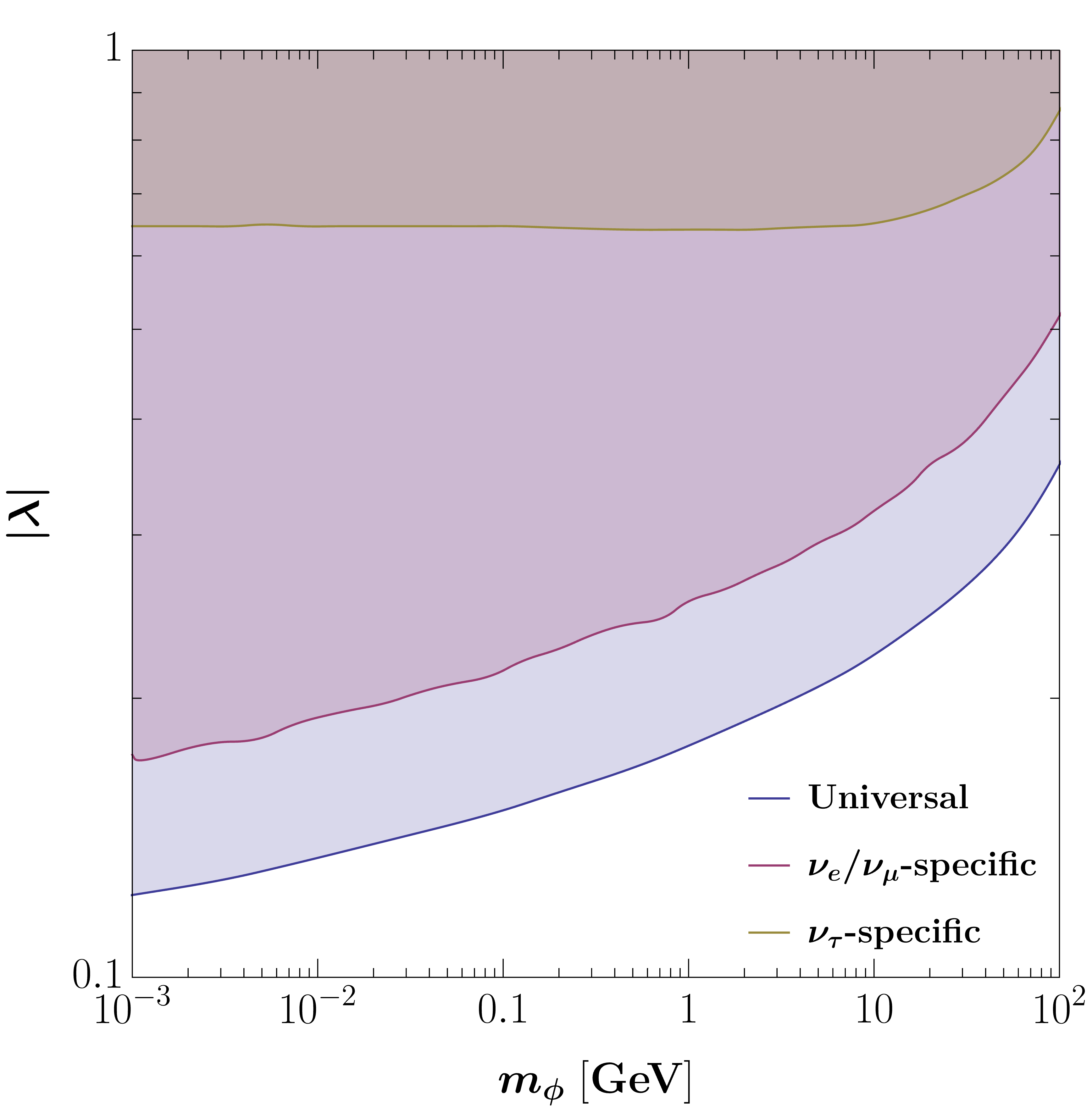}
    \includegraphics[width=0.48\linewidth]{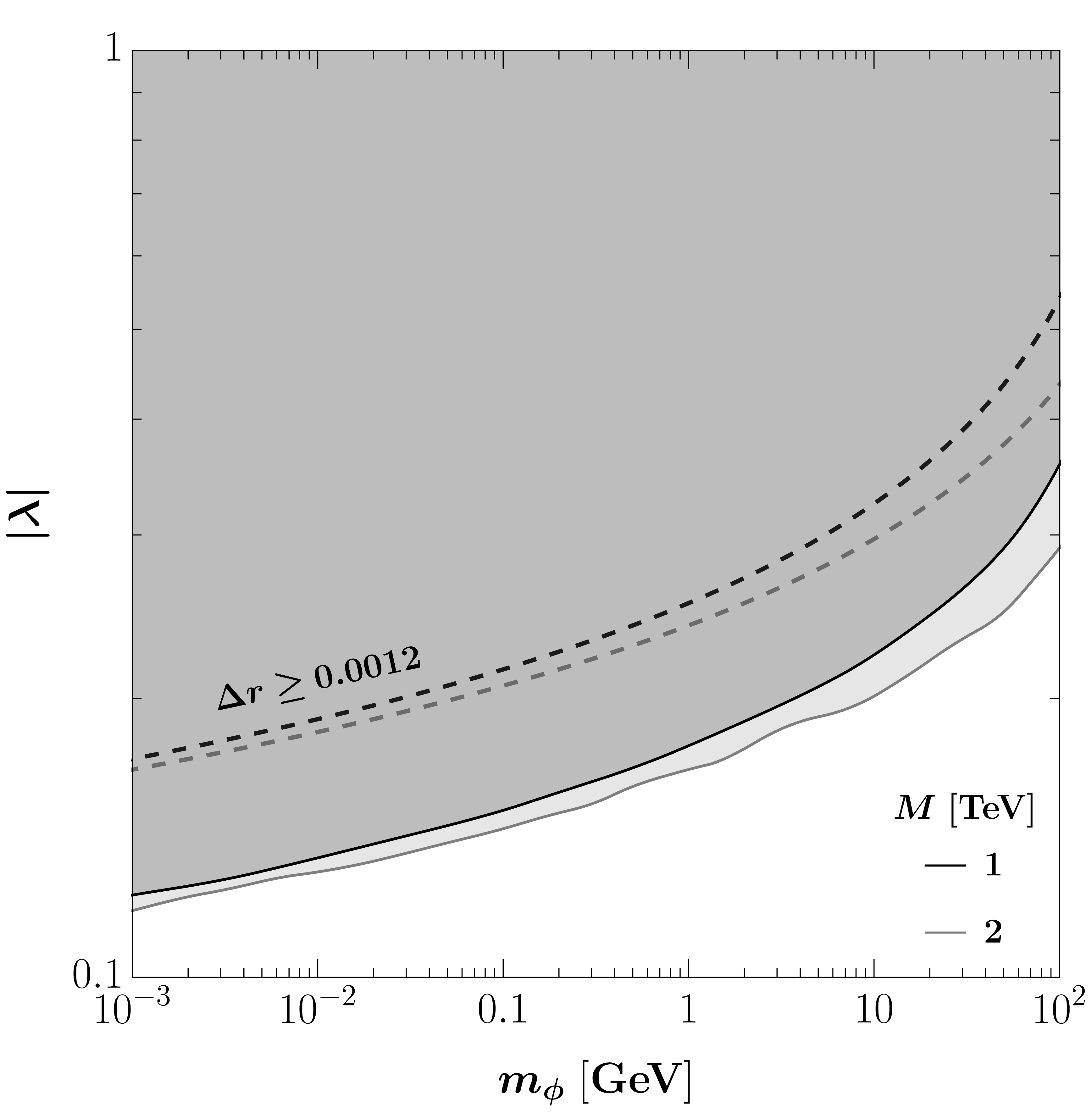}
    \caption{Precision electroweak constraints on the mass and coupling of the neutrinophilic scalar mediating neutrino self-interaction. The shaded regions are excluded at the $2\sigma$ level. \textit{Left:} Constraints for the flavor-universal and flavor-specific cases with $M = 1~{\rm TeV}$; \textit{Right:} Flavour-universal case for two choices of $M$. Regions above the dashed curves denote the exclusion by requiring $|{\Delta r^{{\rm SI}\nu}}|<0.0012$.}
    \label{fig:nsi}
\end{figure}

Let us look into the flavour-universal case of neutrino self-interaction, with 
\begin{align}
\lambda \equiv \lambda_{ee}=\lambda_{\mu\mu}=\lambda_{\tau\tau} \ ,
\end{align}
which has been the most popular benchmark model used for cosmological analyses~\cite{Kreisch:2019yzn, He:2023oke, Camarena:2023cku, Racco:2024lbu}.
In this case, the coupling receives the strongest constraint among the flavor scenarios we consider.
We perform a 2D scan in the $|\lambda|$ versus $m_\phi$ parameter space. The global fit excludes the blue shaded region in the left panel of Fig.~\ref{fig:nsi}, also shown as the gray shaded region in the lower-right panel of Fig.~\ref{fig:res}.
The exclusion is at $2\sigma$ confidence level ($\Delta \chi^2 \leq 6.18$ for 2 new physics parameters) with respect to the best-fit point in the model.
Remarkably, for $m_\phi$ below GeV scale, the coupling $|\lambda|$ is constrained to be less than $\lesssim 0.2$.
%This is a new result. 
For $m_\phi$ above $\sim300$\,MeV, precision electroweak fit gives the leading constraint on $|\lambda|$, and it is much stronger than previously derived from $Z$-boson invisible width alone~\cite{Foroughi-Abari:2025mhj}. 

The right panel of Fig.~\ref{fig:nsi} shows the upper bound on $\lambda$ for two choices of the UV physics scale $M$, corresponding to the black and gray curves, respectively.
The $M$ dependence is logarithmic and mild.
We also use the dashed curves to show the constraint using Eq.~\eqref{eq:deltar} which assumes neutrino self-interaction only affects the $\Delta r^{\rm SI\nu}$ parameter. 
The constraint is slightly weaker which implies that the direct contributions to $Z, W$ boson partial decay widths, albeit less important to the global fit than $\Delta r^{\rm SI\nu}$, are not negligible.

\begin{figure}[!t]
    \centering
    \includegraphics[width=0.9\linewidth]{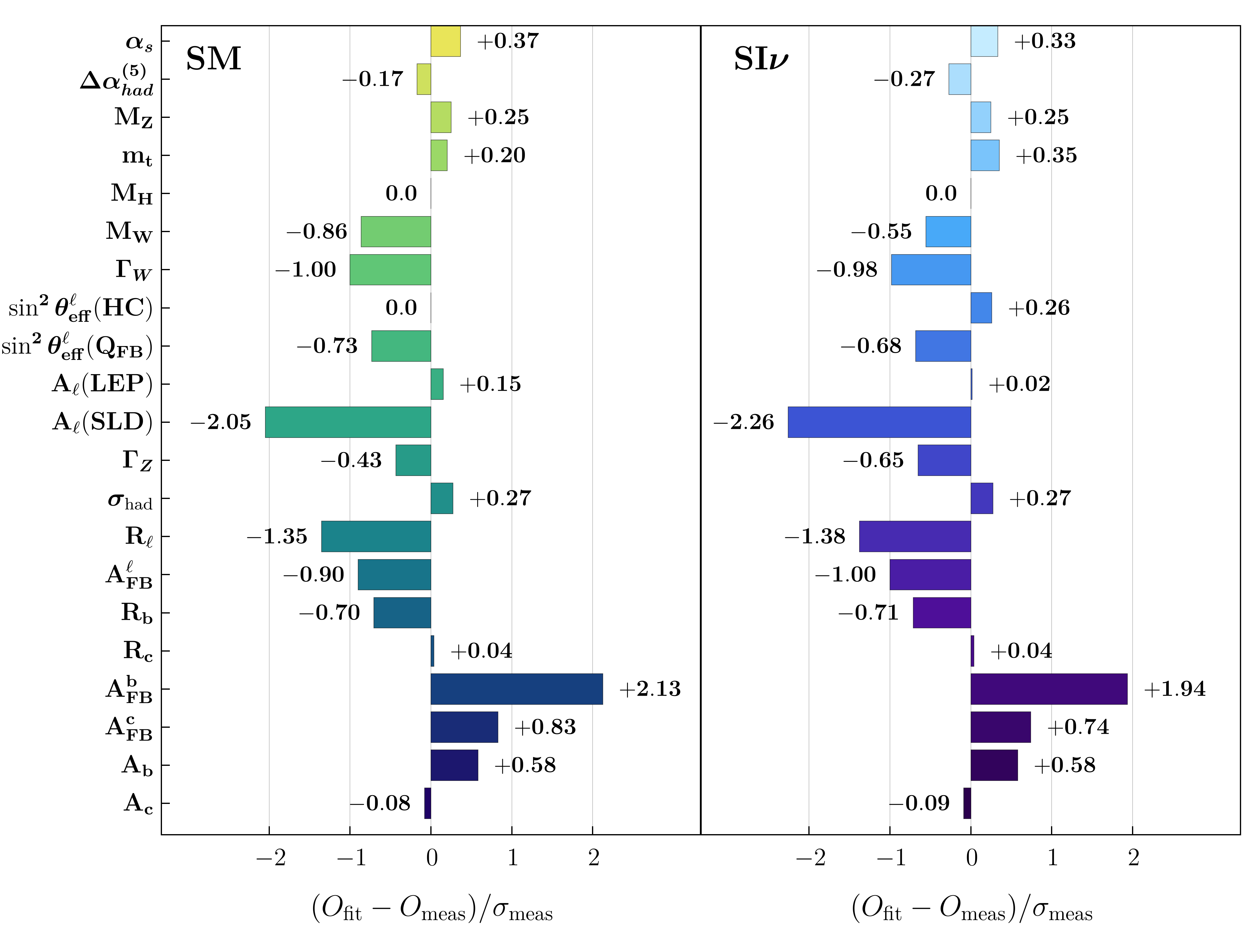}
    \caption{Pull values representing the comparison between the experimental value and fitted value for the SM (\textit{left}) and the universal SI$\nu$ case (\textit{right}).}
    \label{fig:pull}
\end{figure}

The best fit values of parameters and observables are given in the fifth column of Table~\ref{tab:fit}.
It corresponds to a Fermi constant $G_F=1.166649\times10^{-5}\,{\rm GeV}^{-2}$, which is higher than the value of $G_\mu$ in Eq.~\eqref{eq:Gmu}, as expected from the discussion in section~\ref{sec:GF}.
With two new parameters ($|\lambda|, m_\phi$), we obtain a $\chi_{\rm min}^2 = 13.60$, slightly lower than the SM case. 
In the presence of non-Gaussian priors ({\it i.e.}, in our scan, the new physics parameters $|\lambda|$ and $m_\phi$ have flat priors in the log space), the $p$-value cannot be determined by evaluating ${\mathcal{P}}(\chi^2_{\rm min},n_{\rm dof})$, and a toy Monte Carlo (MC) analysis is used. We find a higher $p$-value ${P} \simeq 54\%$ than the SM case, which implies a better goodness of fit. The improvement is rather small, thus by no means shall we claim that the self-interacting neutrino scenario is superior to the SM in the light of electroweak precision tests. A fair takeaway is, at the best fit point, adding novel neutrino self-interaction to the SM does not make the global fit worse.
The comparison of pull values in the SM and the SI$\nu$ scenario are shown in Fig.~\ref{fig:pull}.     
In both cases, the tension with $A_\ell$(SLD) remains above $2\sigma$, while the tension in $A_{\rm FB}^b$ is slightly relaxed in the SI$\nu$ case.

\subsubsection{Flavor-Specific Cases}

For flavor-specific cases, we turn on one coupling $\lambda \equiv \lambda_{\alpha\alpha}$ ($\alpha=e, \mu,\tau$) at a time. 
We perform similar analyses as the flavor-universal case and the corresponding upper bound on $\lambda$ are shown by the purple and brown excluded regions in the left panel of Fig.~\ref{fig:nsi}. They are also shown Fig.~\ref{fig:res}.

For $\nu_e$/$\nu_\mu$-specific case, the $\Delta r^{\rm SI\nu}$ parameter is smaller by a factor of $1/2$ than its counterpart in the flavor-universal case, because in Eq.~\eqref{eq:defDeltar}, $\lambda_{ee}$ and $\lambda_{\mu\mu}$ contribute democratically.
Meanwhile, the direct shift to $W/Z$ to neutrino decay widths is smaller by a factor of $1/3$ because in Eqs.~\eqref{eq:shiftZ2body}, \eqref{eq:shiftZ3body}, \eqref{eq:shiftW2body} and \eqref{eq:shiftW3body}, all the three couplings contribute democratically.
These suppression factors lead to a slightly weaker limit on $|\lambda|$ from the electroweak fit than the flavor-universal case.
Our global analysis finds $\cm \simeq 13.63; \mathcal{P} \simeq 52\%$ with the best fit value of $|\lambda_{ee}|,|\lambda_{\mu\mu}| \simeq 0.09$. 

For $\nu_\tau$-specific case, $\Delta r^{\rm SI\nu}=0$ and there is no shift to the Fermi constant. The only impact of neutrino self-interaction occurs through the direct shift to $W/Z$ to neutrino decay widths. From Table~\ref{tab:fit}, one can see that the $W$ width is much less precisely measured than the $Z$ width, by a factor of $\sim10$, thus the shift to $\Gamma_Z$ carries a much higher weight in the global fit. 
The upper limit on $|\lambda_{\tau\tau}|$ shown in Fig.~\ref{fig:nsi} (brown shaded region) is similar to the invisible $Z$ width constraint found before~\cite{Foroughi-Abari:2025mhj}.
The global fit renders a best-fit value of $|\lambda_{\tau\tau}| \simeq 0.45$, $\cm = 13.54$ and $\mathcal{P} = 54\%$.

As an important message from Fig.~\ref{fig:nsi}, the upper limit on $|\lambda_{\tau\tau}|$ is much weaker than those on $|\lambda_{ee}|, |\lambda_{\mu\mu}|$, and $|\lambda|$ in the flavor universal case.
This comparison reveals the important role of the $\Delta r^{\rm SI\nu}$ parameter in setting the stronger limit on the neutrinophilic coupling in the latter cases.

%%%%%%%%%%%%%%%%%%%%%%%%%%%%%%%%%%%%%%%%%%%%%%%%%%%%
\section{Complementary Constraints}
\label{sec:lowenergy}
%%%%%%%%%%%%%%%%%%%%%%%%%%%%%%%%%%%%%%%%%%%%%%%%%%%%

In this section, we review the other constraints on the neutrinophilic scalar parameter space that is complementary to the precision electroweak constraint derived above. As shown in Fig.~\ref{fig:res}, some of these constraints are flavor dependent and more important for small $\phi$ masses.

First of all, with a coupling $\lambda\gtrsim 10^{-8}$, the $\phi$ particles will thermalize with the SM neutrinos in the early universe. This sets a lower bound on $m_\phi$ to avoid excessive contribution to the radiation energy density ($\Delta N_{\rm eff}$) around the time of the big-bang nucleosynthesis (BBN)~\cite{Nollett:2013pwa}. For $\phi$ as a complex scalar, the mass lower bound is about 6\,MeV~\cite{Li:2023puz}. The magenta regions are excluded in Fig.~\ref{fig:res}.

A sufficiently light neutrinophilic scalar $\phi$ can be radiated by the final state neutrino in the regular leptonic decay of charged mesons $\mathfrak{m}^+ = \pi^+, K^+$, leading to exotic three-body decay channels. 
These constraints have been explored and revisited in the literature~\cite{Barger:1981vd, Berryman:2018ogk, Blinov:2019gcj, Brdar:2020nbj, Dev:2024ygx, deLima:2026fsy}.
Experimental upper bounds on $\mathfrak{m}^+ \to \ell_\alpha^+ \nu_\alpha X$ (here $X=\phi$ and $\alpha = e, \mu$) have recently been set by PIENU~\cite{PIENU:2021clt}, NA62~\cite{NA62:2021bji} (see also~\cite{Heintze:1977kk}).
With a bit more details, the branching ratio of $\mathfrak{m}^+ \to \ell_\alpha^+ \bar\nu_\alpha \phi$ is given by~\cite{deLima:2026fsy}
\begin{equation} \label{eq:chargedmeson}
\begin{split}
\frac{{\rm Br}({\mathfrak{m}^+ \rightarrow \ell_\alpha^+ \bar\nu_\alpha \phi})}{{\rm Br}({\mathfrak{m}^+ \rightarrow \ell_\alpha^+ \nu_\alpha})} 
= \int_{2 \sqrt{\delta_\ell}}^{1 +\delta_\ell - \delta_{\phi}} d x_\ell
\frac{|\lambda_{\alpha\alpha}|^2 \sqrt{x_\ell^2-4 \delta _\ell} \left[\left(x_\ell-1\right) x_\ell-2 \delta_\ell\right] \left(\delta _{\phi}-\delta_\ell+x_\ell-1\right)^2}{32 \pi^2 \delta_\ell \left(\delta_\ell-1\right)^2 \left(-\delta_\ell+x_\ell-1\right)^3} \ ,
\end{split}
\end{equation} 
where $\delta _\ell={m_{\ell_\alpha}^2}/{m_{\mathfrak{m}}^2}$, $\delta _\phi={m_\phi^2}/{m_{\mathfrak{m}}^2}$, and $x_\ell=2 E_\ell/m_{\mathfrak{m}}$ with $E_\ell$ being the final state charged lepton energy in the rest frame of $\mathfrak{m}$. The $x_\ell$ integral can be done numerically.
${\rm Br}({\mathfrak{m}^+ \rightarrow \ell_\alpha^+ \nu_\alpha})$ is the regular leptonic decay branching ratio measured experimentally.
In Eq.~\eqref{eq:chargedmeson}, the dependence on meson decay constants ($f_\pi, f_K$, whose values are subject to large theoretical uncertainties~\cite{PDGdecayconstants}) and the CKM matrix elements ($V_{ud}$, $V_{us}$), nicely cancel away, allowing robust upper bound on $\lambda$ to be set as a function of $m_\phi$.
They exclude the orange and blue regions in Fig.~\ref{fig:res}.
The NA62 bound was set for $m_\phi$ above 10\,MeV~\cite{NA62:2021bji}; We simply extrapolate it down to 6 MeV where the BBN constraint takes over.

The above meson decay constraints apply for $\lambda_{ee}, \lambda_{\mu\mu}$ and the flavor-universal cases, but not for $\lambda_{\tau\tau}$. In the latter case, there are two useful constraints other than BBN.
One is derived based on the IceCube measurement of the spectrum of ultra-high energy astrophysical neutrinos and excludes their strong resonant scattering with the cosmic neutrino background~\cite{Esteban:2021tub}.
This excludes the brown region in the lower-left panel of Fig.~\ref{fig:res}.
The other constraint arises from the breaking of lepton flavor universality in $Z$-mediated neutrino-matter interaction which affects the global fit to neutrino oscillation data~\cite{Foroughi-Abari:2025mhj, Coloma:2023ixt} and excludes the red shaded region in lower left panel of Fig.~\ref{fig:res}.

The precision electroweak and neutrino oscillation constraints have a mild, logarithmic dependence on the heavy mass scale $M$. In Fig.~\ref{fig:res}, we set $M=1\,$TeV.  
In this case, the precision electroweak fit explored in this work gives the leading constraint for $m_\phi\gtrsim 300\,$MeV in the case of $\nu_e, \nu_\mu$-specific and the flavor universal neutrino self-interaction, and for $m_\phi\gtrsim 1\,$GeV in the $\nu_\tau$-specific case.

%%%%%%%%%%%%%%%%%%%%%%%%%%%%%%%%%%%%%
\section{Precision Electroweak Constraint on a UV Completion}
\label{sec:UVcompletion}
%%%%%%%%%%%%%%%%%%%%%%%%%%%%%%%%%%%%%

In this section, we discuss a concrete UV completion that can accommodate the light neutrinophilic $\phi$ in Eq.~\eqref{eq:lagphi}.
This is by extending the SM with a complex scalar $T$ that transforms as a $SU(2)_L$ triplet and carries hypercharge $-2$,
\begin{align}
    T = \begin{pmatrix}
        T^-/\sqrt{2} & T^0\\
        T^{--} & -T^-/\sqrt{2}
    \end{pmatrix}\,.
\end{align}
The new physics part of the Lagrangian is given by
\begin{equation}\label{eq:lagT}
    \begin{split}
    \mathcal{L} &= \text{Tr}[(D^\mu T)^\dagger(D_\mu T)] - M^2 \text{Tr}[T^\dagger T] + \partial^\mu \phi^\dagger \partial_\mu \phi - \mu_\phi^2 |\phi|^2 \\
    &\quad - \sum_{\alpha,\beta = e,\mu,\tau} \frac{y_{\alpha \beta}}{2}\bar{L}_\alpha T i\sigma_2 L^c_\beta - y' H^T i\sigma_2 T H \phi^\ast  + \hc\ .
\end{split}
\end{equation}
The covariant derivative of $T$ is given by $D_\mu T \equiv \del_\mu T +i(g/2)W_\mu^a [\sigma_a, T] -ig' B_\mu T$~\cite{Kriewald:2024cgr}, 
where $W_\mu$ and $B_\mu$ are the $SU(2)_L$ and $U(1)_Y$ gauge bosons, respectively. 
The neutral component of the triplet $T^0$ only couples to neutrinos via the Yukawa interaction in the second line.
However, it also carries electroweak charge and couples to the $Z$ boson and cannot have a mass well below the weak scale.
In order for the model to accommodate a light mediator for neutrino self-interaction, we also introduce a gauge-singlet complex scalar $\phi$.

The presence of $T$ is similar to Type-II seesaw mechanism for neutrino masses~\cite{Schechter:1980gr,Lazarides:1980nt,Cheng:1980qt,Mohapatra:1980yp,Schechter:1981cv}, however, we assume that neither $T$ or $\phi$ gets a VEV, keeping the focus only on neutrino self-interactions. 
The nonzero neutrino mass will be explained by other sources, for example, Dirac masses as discussed in~\cite{Berryman:2018ogk}. Note that the entire Lagrangian in Eq.~\eqref{eq:lagT} has a $U(1)_{B-L}$ global symmetry under which both $T$ and $\phi$ carry $B-L$ charge $-2$. Thanks to this symmetry, a trilinear term involving the SM Higgs doublet and just the triplet $T$ found in the Type-II seesaw model is forbidden here, thereby avoiding any mixing between $T^0$ and the SM Higgs. A mixing between the electric neutral scalars $\phi$ and $T^0$ is generated after spontaneous electroweak symmetry breaking, i.e., $\braket{H} = (0,~v/\sqrt{2})^T$ with $v=246~\GeV$. The  mass-mixing matrix in the basis $\{\phi, T^0\}$ can be written as 
\begin{align}\label{eq:massmix}
    \mathcal{M}^2 = \begin{pmatrix}
        \mu_\phi^2 & y'v^2/2\\
        y'^*v^2/2 & M^2
    \end{pmatrix}\,,
\end{align}
where $\mu_\phi^2$ corresponds to the mass parameter for $\phi$. We are free to redefine the $\phi$ field to absorb the phase of the $y'$ parameter. The mass eigenstates $\hat{\phi}$ and $\hat{T}^0$ can be obtained upon diagonalization of Eq.~\eqref{eq:massmix}
\begin{align}
    \hat{\mathbf{m}}^2 \equiv \text{diag}(m_{\hat{\phi}}^2,m_{\hat{T}^0}^2) = \mathbf{R}^T \mathcal{M}^2\, \mathbf{R}\,,
\end{align}
where $\mathbf{R}$ is the diagonalizing matrix characterized by a Euler angle $\theta$, which satisfies the relation
\begin{align}
\tan(2\theta) = \frac{|y'|v^2}{M^2 - \mu_\phi^2} \ .
\end{align}
We will work in the large $M$ limit where
\begin{align}\label{eq:thetasmall}
    \sin\theta \simeq \frac{|{y'}|v^2}{2 M^2} \simeq 0.03 |y'| \left( \frac{1\,\rm TeV}{M} \right)^2 \ .
\end{align}
and the scalar masses eigenvalues are
\begin{align}
    m_{\hat{\phi}}^2 = \mu_\phi^2 - \frac{|{y'}|^2 v^4}{4M^2}\,,\quad m_{\hat{T}^0}^2 \simeq M^2 + \frac{|{y'}|^2 v^4}{4M^2}\,.
\end{align}
For simplicity, we will ignore the hat over the physical fields for the rest of the paper. 
It is interesting to note that the above scalar mixing lifts the mass of $T^0$, allowing it to be heavier than the $T^-, T^{--}$ partners from the triplet. 
However, such a mass splitting, $\delta m=\sqrt{M^2 + {|{y'}|^2 v^4}/{(4M^2)}} - M$, is rather small. Even for order one $|y'|$, $\delta m\lesssim 1\,$GeV, which is not of phenomenological relevance for this study. We neglect it hereafter.~\footnote{Another source of mass splitting arises from electroweak loop corrections~\cite{Cirelli:2005uq}, which is also small and neglected.} On the other hand, the triplet component masses can be made different by a scalar potential term, $\beta H^\dagger T T^\dagger H$~\cite{Melfo:2011nx}
\begin{align}\label{eq:masssplitting}
    m_{T^-}^2 - m_{T^{--}}^2 = m_{T^0}^2 - m_{T^-}^2 = \frac{1}{4} \beta v^2 \ .
\end{align}
The mass squares receive equal splitting. Without loss of generality, we write
\begin{align}\label{eq:tripletcomponentmasses}
    m_{T^-} = M, \quad m_{T^0} = M + \Delta M, \quad m_{T^{--}} = \sqrt{M^2 - 2 M \Delta M - (\Delta M)^2} \ .
\end{align}

The $\phi-T^0$ mixing allows $\phi$ to participate in interactions with the SM sector. Essentially, for every interacting vertex involving a $T^0$ in the $\theta=0$ limit, it should now be replaced by the linear combination $(T^0 \cos\theta + \phi \sin\theta)$.
As a result, $\phi$ inherits all the $T^0$ interactions but with couplings suppressed by a universal factor $\sin\theta$. In particular, the Yukawa coupling in Eq.~\eqref{eq:lagphi} is related to the fundamental coupling $y$ via~\footnote{The doubly-charged component $T^{--}$ only couples to a pair of charged leptons. As mentioned in the paragraph below Eq.~\eqref{eq:f1}, we assume $y$ to be flavor diagonal, $y_{\alpha\beta} = 0$ for $\alpha\neq \beta$. This choice forbids the dangerous lepton flavor changing decays mediated by $T^{--}$~\cite{Barrie:2022ake}.}
\begin{align}\label{eq:lambday}
    \lambda_{\alpha \beta} \simeq y_{\alpha \beta} \sin\theta \,.
\end{align}
$\phi$ also obtains $\sin\theta$-suppressed gauge couplings from $T^0$.
Meanwhile, the Yukawa and gauge couplings of the physical $T^0$ after the mixing are rescaled by a factor of $\cos\theta$.

\subsection{Tree-level $Z\to \phi \phi^*$ Decay}\label{sec:Zphiphi*}

An immediate consequence of the $\phi$-$T^0$ mixing is a new tree-level decay mode of the $Z$ boson into $\phi\phi^*$, which is kinematically allowed for $m_\phi<m_Z/2$. The decay rate is
\begin{align}\label{eq:Zphiphi}
    \Gamma_{Z\r \phi\phi^\ast} = \frac{\sqrt{2}\gf m_Z^3 \sin^4\theta}{12\pi}\left(1-\frac{4m_\phi^2}{m_Z^2}\right)^{3/2} \ ,
\end{align}
and the final state $\phi$'s will further decay into neutrinos.
Although this is a $\sin^4\theta$ order quantity, we find it to be numerically important.
In~\cite{Foroughi-Abari:2025mhj}, an upper bound $|\sin\theta|<0.3$ was obtained by requiring $\Gamma_{Z\r \phi\phi^\ast}<3\,$MeV~\cite{ParticleDataGroup:2024cfk}. In the global analysis of this work, we will add this decay width to the total $Z$ width in Eq.~\eqref{eq:GammaZ}. In contrast, we neglect the three-body decay mode $Z\to \nu\nu\phi^*$ which can take place via off-shell $T^0$ but is phase space suppressed and subdominant.

\subsection{Structure of Radiative Corrections to $\delta g_Z$, $\delta g_W$}
\label{sec:structureofRC}

Moving on to loops,
we first clarify the structure of radiative corrections to the neutrino gauge couplings $\delta g_Z$, $\delta g_W$ in this UV complete model and their relationship with those derived in the simplified model in Eq.~\eqref{eq:dgzdgw}.
In the presence of the mixing, both $\phi$ and $T$ contribute at one-loop level, through the gauge vertex and self-energy diagrams.
The finiteness of $\delta g_Z$, $\delta g_W$ is guaranteed by gauge invariance and the Ward identity.
We leave the discussion of gauge boson vacuum polarization diagrams in section~\ref{sec:oblique}, which contribute as the electroweak oblique parameters at energy scales well below the triplet mass.

In general, the correction to $\delta g_V$ ($V=Z, W$) can be written as
\begin{align}\label{eq:gV}
    \delta g_V = (\delta g_V)_{\phi} + (\delta g_V)_{T} \ ,
\end{align}
where $(\delta g_V)_{\phi}$ receives contributions from loop diagrams that contain at least one $\phi$. It occurs at $\sin^2\theta$ order or higher.
In contrast, $(\delta g_V)_{T}$ is from diagrams with only the triplet $T$ in the loop.

It is useful to reorganize Eq.~\eqref{eq:gV} as the following
\begin{equation}\label{eq:gV2}
\begin{split}
    \delta g_V &= (\delta g_V)_{\phi} + \left[ (\delta g_V)_{T} - (\delta g_V)_{T,\, \theta=0} \rule{0mm}{4mm}\right] \\
    & \quad + (\delta g_V)_{T,\, \theta=0} \ .
\end{split}
\end{equation}
The quantity in the square bracket is nonzero due to the different Yukawa and gauge couplings of $T$ in the $\theta\neq0$ and $\theta=0$ cases. It is also of $\sin^2\theta$ order or higher.

The second line of Eq.~\eqref{eq:gV2} represents the contribution of $T$ in the absence of $\phi-T^0$ mixing whose finiteness is another consequence of gauge invariance. As a result, both lines of Eq.~\eqref{eq:gV2} are free from UV divergences.

The first line of Eq.~\eqref{eq:gV2} has already been carefully calculated in~\cite{Foroughi-Abari:2025mhj} and the results are given by Eq.~\eqref{eq:dgzdgw} with $c_Z=-11/2$ and $c_W=-2$. Up to a loop factor, this contribution is of order
\begin{align}\label{eq:line1}
\lambda^2 \ln\frac{M^2}{m_\phi^2} \sim \frac{y^2 y'^2 v^4}{M^4} \ln\frac{M^2}{m_\phi^2} \ ,
\end{align}
where in the second step we used Eqs.~\eqref{eq:thetasmall} and \eqref{eq:lambday}. 
In comparison, the second line of Eq.~\eqref{eq:gV2} corresponds to the radiative correction from $T$ in the absence of $\phi$ field.
It will be calculated in the next section and the contributions are of order
\begin{align}\label{eq:line2}
 \frac{y^2 q^2}{M^2} \ ,
\end{align}
where the logarithmic factor is absent.

As we already see in Figs.~\ref{fig:res} and~\ref{fig:nsi}, the precision electroweak constraints are sensitive to $\lambda \gtrsim 0.1$. This corresponds to the triplet mass $M$ around TeV scale.
For sufficiently small $m_\phi$, the logarithmic factor is large allowing the first line of Eq.~\eqref{eq:gV2} to be important, which has been explored in~\cite{Foroughi-Abari:2025mhj}.
On the other hand, the prefactor in front of the log is suppressed compared to Eq.~\eqref{eq:line2} by an extra $M^2$. 
This implies that the heavy $T$ contribution cannot be simply ignored for $Z$-pole physics where $q^2=m_Z^2$.
The competition is between $\sin^2\theta\cdot\ln(M^2/m_\phi^2)$ and $(m_Z/M)^2$.

\subsection{Heavy $T$ contribution to $\delta g_Z$, $\delta g_W$ and $Z\ell^+\ell^-$ Couplings}
\label{sec:gaugevertex}

In this work, we complete the calculation of~\cite{Foroughi-Abari:2025mhj} by including the second line of Eq.~\eqref{eq:gV2}, {\it i.e.}, radiative correction to gauge couplings from the heavy triplet $T$. 
As already discussed, this contribution occurs in $\theta=0$ limit.

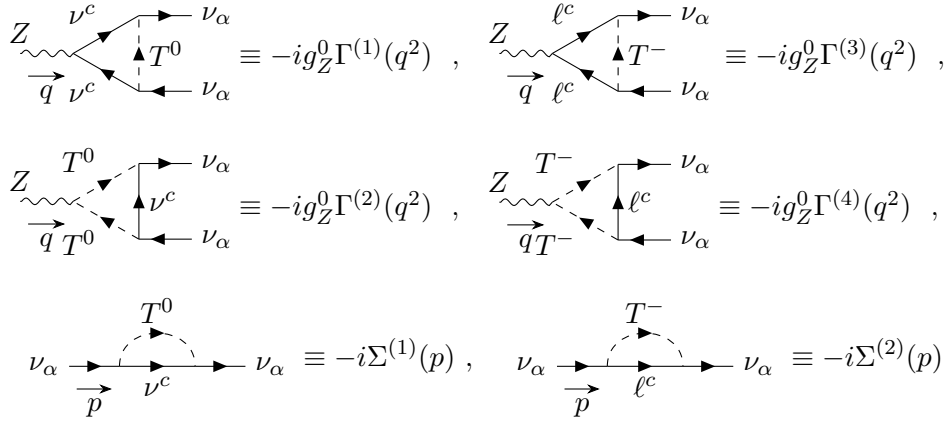
\begin{figure}[h]
    \centering
    \begin{align}
&\begin{tikzpicture}
\begin{feynman}
\vertex [label=above:\(Z\)] (a) at (0,0);
\vertex (b) at (0.7,0);
\vertex (c) at (1.57,0.5);
\vertex (d) at (1.57,-0.5);
\vertex [label=right:\(\nu_\alpha\)] (e) at (2.27,0.5);
\vertex [label=right:\(\nu_\alpha\)] (f) at (2.27,-0.5);
\feynmandiagram [inline=(a.base)] {
(a)  -- [photon, momentum'=\(q\)] (b),
(b) -- [fermion, edge label=\(\nu^c\)] (c),
(d) -- [charged scalar, edge label'=\(T^0\)] (c),
(d) -- [fermion, edge label=\(\nu^c\)] (b),
(c) -- [fermion] (e),
(d) -- [anti fermion] (f),
};
\end{feynman}
\end{tikzpicture} 
\hspace{0.5cm}
\raisebox{0.65cm}{
$\equiv -i g_{Z}^0 \Gamma^{(1)}(q^2)$ \ ,
} \quad 
\begin{tikzpicture}
\begin{feynman}
\vertex [label=above:\(Z\)] (a) at (0,0);
\vertex (b) at (0.7,0);
\vertex (c) at (1.57,0.5);
\vertex (d) at (1.57,-0.5);
\vertex [label=right:\(\nu_\alpha\)] (e) at (2.27,0.5);
\vertex [label=right:\(\nu_\alpha\)] (f) at (2.27,-0.5);
\feynmandiagram [inline=(a.base)] {
(a)  -- [photon, momentum'=\(q\)] (b),
(b) -- [fermion, edge label=\(\ell^c\)] (c),
(d) -- [charged scalar, edge label'=\(T^-\)] (c),
(d) -- [fermion, edge label=\(\ell^c\)] (b),
(c) -- [fermion] (e),
(d) -- [anti fermion] (f),
};
\end{feynman}
\end{tikzpicture} 
\hspace{0.5cm}
\raisebox{0.65cm}{
$\equiv -i g_{Z}^0 \Gamma^{(3)}(q^2)$ \ ,
}
 \nonumber \\
&\begin{tikzpicture}
\begin{feynman}
\vertex [label=above:\(Z\)] (a) at (0,0);
\vertex (b) at (0.7,0);
\vertex (c) at (1.57,0.5);
\vertex (d) at (1.57,-0.5);
\vertex [label=right:\(\nu_\alpha\)] (e) at (2.27,0.5);
\vertex [label=right:\(\nu_\alpha\)] (f) at (2.27,-0.5);
\feynmandiagram [inline=(a.base)] {
(a)  -- [photon, momentum'=\(q\)] (b),
(b) -- [charged scalar, edge label=\(T^0\)] (c),
(d) -- [fermion, edge label'=\(\nu^c\)] (c),
(d) -- [charged scalar, edge label=\(T^0\)] (b),
(c) -- [fermion] (e),
(d) -- [anti fermion] (f),
};
\end{feynman}
\end{tikzpicture} 
\hspace{0.5cm}
\raisebox{0.65cm}{
$\equiv -i g_{Z}^0 \Gamma^{(2)}(q^2)$ \ ,
}  \quad
\begin{tikzpicture}
\begin{feynman}
\vertex [label=above:\(Z\)] (a) at (0,0);
\vertex (b) at (0.7,0);
\vertex (c) at (1.57,0.5);
\vertex (d) at (1.57,-0.5);
\vertex [label=right:\(\nu_\alpha\)] (e) at (2.27,0.5);
\vertex [label=right:\(\nu_\alpha\)] (f) at (2.27,-0.5);
\feynmandiagram [inline=(a.base)] {
(a)  -- [photon, momentum'=\(q\)] (b),
(b) -- [charged scalar, edge label=\(T^-\)] (c),
(d) -- [fermion, edge label'=\(\ell^c\)] (c),
(d) -- [charged scalar, edge label=\(T^-\)] (b),
(c) -- [fermion] (e),
(d) -- [anti fermion] (f),
};
\end{feynman}
\end{tikzpicture} 
\hspace{0.5cm}
\raisebox{0.65cm}{
$\equiv -i g_{Z}^0 \Gamma^{(4)}(q^2)$ \ ,
} 
\nonumber \\
&\feynmandiagram [inline=(a.base), horizontal=a to d] {
  i1 [particle=\(\nu_\alpha\)] -- [fermion, momentum'=\(p\)] a -- [fermion, edge label'=\(\nu^c\)] b -- [fermion] i2 [particle=\(\nu_\alpha\)],
  a -- [charged scalar, half left, looseness=1.5, edge label=\(T^0\)] b,
}; 
\equiv -i\Sigma^{(1)}(p) \ , \quad 
\feynmandiagram [inline=(a.base), horizontal=a to d] {
  i1 [particle=\(\nu_\alpha\)] -- [fermion, momentum'=\(p\)] a -- [fermion, edge label'=\(\ell^c\)] b -- [fermion] i2 [particle=\(\nu_\alpha\)],
  a -- [charged scalar, half left, looseness=1.5, edge label=\(T^-\)] b,
}; 
\equiv -i\Sigma^{(2)}(p) \nonumber
\end{align} 
    \caption{One-loop vertex and self-energy corrections to the $Z \nu \bar{\nu}$ vertex due to the heavy scalars $T^0$ and $T^-$.}
    \label{fig:radnunuT}
\end{figure}

We first calculate the correction to the $Z\nu\bar\nu$ coupling, $g_Z$ defined in Eq.~\eqref{eq:EWL}. The relevant Feynman diagrams are shown in Fig.~\ref{fig:radnunuT}.
Assuming flavor diagonal Yukawa couplings, $y_{\alpha\beta} = y_{\alpha\alpha} \delta_{\alpha\beta}$, the corresponding loop functions are
\begin{align}\label{eq:GammaZnunu}
\Gamma^{(1)} = & -\frac{|y_{\alpha\alpha}|^2}{32\pi^2} 
\left[ \frac{2}{\varepsilon} - 1 + f_1 (q^2, M, \mu) \right]\ , \quad \Gamma^{(3)} = \frac{1}{2} (-1+2\sin^2\theta_W) \Gamma^{(1)} \ , \nonumber\\
\Gamma^{(2)} =& \frac{|y_{\alpha\alpha}|^2}{16\pi^2} 
\left[ \frac{2}{\varepsilon} + f_2 (q^2, M, \mu)  \right] \ , \quad \Gamma^{(4)} = \frac{1}{2} \sin^2\theta_W \Gamma^{(2)} \ ,\nonumber\\
\Sigma^{(1)} =& -\frac{|y_{\alpha\alpha}|^2}{32\pi^2} \left( \frac{2}{\varepsilon} + \frac{1}{2} + \log \frac{\mu^2}{M^2}\right) \cancel{p} \ , \quad \Sigma^{(2)} = \frac{1}{2} \Sigma^{(1)} \ ,
\end{align}
where $\mu$ is the renormalization scale and the loop function $f_1$ has been defined in Eq.~\eqref{eq:f1} and $f_2$ is given by
\begin{align}
  f_2 (q^2, M, \mu) =  2\int_0^1 dx \int_0^{1-x}dy \ln \frac{\mu^2}{(x+y)M^2-xy q^2}
\end{align}
Their correction to $g_Z$ is
\begin{eqnarray}\label{eq:deltagZfromT}
\left.\frac{(\delta g_{Z})_{\alpha\alpha}}{g_{Z}^0}\right|_{T,\, \theta=0} = \sum_{i=1}^4 \Gamma^{(i)} + \sum_{j=1}^2\frac{\partial \Sigma^{(j)}}{\partial \cancel{p}} \ .
\end{eqnarray}
All the $1/\varepsilon$ divergences and the $\ln\mu$ dependence cancel away in the sum, as expected. In the large $M$ limit, Eq.~\eqref{eq:deltagZfromT} vanishes as $q^2/M^2$, which verifies the decoupling theorem.
We add the heavy $T$ contribution as part of $\delta g_Z$ in Eq.~\eqref{eq:shiftZ2body}.

The heavy triplet loop can also give correction to the coupling between the $Z$ boson and left-handed charged leptons. The neutral-current interacting Lagrangian for charged leptons is
\begin{equation}\label{eq:ellNC}
\begin{split}
\mathcal{L} & = - g_Z^0 \bar{\ell}_\alpha \gamma^\mu \left( g_{V, \ell_\alpha} - g_{A,\ell_\alpha} \gamma_5 \right)\ell_\alpha Z_\mu \\
&= - g_Z^0 \bar{\ell}_\alpha \gamma^\mu \left[ - \sqrt{\rho_Z^{\ell_\alpha}} \mathbb{P}_L + 2 \sqrt{\rho_Z^{\ell_\alpha}} \kappa_Z^{\ell_\alpha} \sin^2\theta_W (\mathbb{P}_L + \mathbb{P}_R) \right]\ell_\alpha Z_\mu \ ,
\end{split}
\end{equation}
where we used Eq.~\eqref{eq:gz} and $\ell_\alpha = e^-, \mu^-, \tau^-$.
Because the right-handed current is not affected by $T$, the radiative correction must keep the product $\sqrt{\rho_Z^{\ell_\alpha}} \kappa_Z^{\ell_\alpha}$ intact, which implies
\begin{align}
     \left(\sqrt{\rho_Z^{\ell_\alpha}} -1 \right)_{T,\, \theta=0} \simeq - \left(\kappa_Z^{\ell_\alpha} -1\right)_{T,\, \theta=0} \ .
\end{align}
Only the first term in the square bracket in Eq.~\eqref{eq:ellNC}, which is proportional to the weak isospin, is affected by $T$ loops. 
The corresponding Feynman diagrams are shown in Fig.~\ref{fig:radllT}.
\begin{figure}[h]
    \centering
    \resizebox{\textwidth}{!}{%
    \begin{minipage}{\textwidth}
    \begin{align}
&\begin{tikzpicture}
\begin{feynman}
\vertex [label=above:\(Z\)] (a) at (0,0);
\vertex (b) at (0.7,0);
\vertex (c) at (1.57,0.5);
\vertex (d) at (1.57,-0.5);
\vertex [label=right:\(\ell_\alpha\)] (e) at (2.27,0.5);
\vertex [label=right:\(\ell_\alpha\)] (f) at (2.27,-0.5);
\feynmandiagram [inline=(a.base)] {
(a)  -- [photon, momentum'=\(q\)] (b),
(b) -- [fermion, edge label=\(\nu^c\)] (c),
(d) -- [charged scalar, edge label'=\(T^-\)] (c),
(d) -- [fermion, edge label=\(\nu^c\)] (b),
(c) -- [fermion] (e),
(d) -- [anti fermion] (f),
};
\end{feynman}
\end{tikzpicture} 
\hspace{0.5cm}
\raisebox{0.65cm}{
$\equiv -i g_{Z}^0 \mathbb{P}_L \Gamma'^{(1)}(q^2)$ \ ,
} \quad 
&&\begin{tikzpicture}
\begin{feynman}
\vertex [label=above:\(Z\)] (a) at (0,0);
\vertex (b) at (0.7,0);
\vertex (c) at (1.57,0.5);
\vertex (d) at (1.57,-0.5);
\vertex [label=right:\(\ell_\alpha\)] (e) at (2.27,0.5);
\vertex [label=right:\(\ell_\alpha\)] (f) at (2.27,-0.5);
\feynmandiagram [inline=(a.base)] {
(a)  -- [photon, momentum'=\(q\)] (b),
(b) -- [fermion, edge label=\(\ell^c\)] (c),
(d) -- [charged scalar, edge label'=\(T^{--}\)] (c),
(d) -- [fermion, edge label=\(\ell^c\)] (b),
(c) -- [fermion] (e),
(d) -- [anti fermion] (f),
};
\end{feynman}
\end{tikzpicture} 
\hspace{0.5cm}
\raisebox{0.65cm}{
$\equiv -i g_{Z}^0 \mathbb{P}_L \Gamma'^{(3)}(q^2)$ \ ,
}
 \nonumber \\
&\begin{tikzpicture}
\begin{feynman}
\vertex [label=above:\(Z\)] (a) at (0,0);
\vertex (b) at (0.7,0);
\vertex (c) at (1.57,0.5);
\vertex (d) at (1.57,-0.5);
\vertex [label=right:\(\ell_\alpha\)] (e) at (2.27,0.5);
\vertex [label=right:\(\ell_\alpha\)] (f) at (2.27,-0.5);
\feynmandiagram [inline=(a.base)] {
(a)  -- [photon, momentum'=\(q\)] (b),
(b) -- [charged scalar, edge label=\(T^-\)] (c),
(d) -- [fermion, edge label'=\(\nu^c\)] (c),
(d) -- [charged scalar, edge label=\(T^-\)] (b),
(c) -- [fermion] (e),
(d) -- [anti fermion] (f),
};
\end{feynman}
\end{tikzpicture} 
\hspace{0.5cm}
\raisebox{0.65cm}{
$\equiv -i g_{Z}^0 \mathbb{P}_L \Gamma'^{(2)}(q^2)$ \ ,
}  \quad
&&\begin{tikzpicture}
\begin{feynman}
\vertex [label=above:\(Z\)] (a) at (0,0);
\vertex (b) at (0.7,0);
\vertex (c) at (1.57,0.5);
\vertex (d) at (1.57,-0.5);
\vertex [label=right:\(\ \ell_\alpha\)] (e) at (2.27,0.5);
\vertex [label=right:\(\ \ell_\alpha\)] (f) at (2.27,-0.5);
\feynmandiagram [inline=(a.base)] {
(a)  -- [photon, momentum'=\(q\ \ \ \ \  \)] (b),
(b) -- [charged scalar, edge label=\(T^{--}\)] (c),
(d) -- [fermion, edge label'=\(\ell^c\)] (c),
(d) -- [charged scalar, edge label=\(T^{--}\)] (b),
(c) -- [fermion] (e),
(d) -- [anti fermion] (f),
};
\end{feynman}
\end{tikzpicture} 
\hspace{0.5cm}
\raisebox{0.65cm}{
$\equiv -i g_{Z}^0 \mathbb{P}_L \Gamma'^{(4)}(q^2)$ \ ,
} 
\nonumber \\
&\feynmandiagram [inline=(a.base), horizontal=a to d] {
  i1 [particle=\(\ell_\alpha\)] -- [fermion, momentum'=\(p\)] a -- [fermion, edge label'=\(\nu^c\)] b -- [fermion] i2 [particle=\(\ell_\alpha\)],
  a -- [charged scalar, half left, looseness=1.5, edge label=\(T^-\)] b,
}; 
\equiv -i\Sigma'^{(1)}(p) \ , \quad
&&\feynmandiagram [inline=(a.base), horizontal=a to d] {
  i1 [particle=\(\ell_\alpha\)] -- [fermion, momentum'=\(p\)] a -- [fermion, edge label'=\(\ell^c\)] b -- [fermion] i2 [particle=\(\ell_\alpha\)],
  a -- [charged scalar, half left, looseness=1.5, edge label=\(T^{--}\)] b,
}; 
\equiv -i\Sigma'^{(2)}(p) \nonumber
\end{align} 
    \end{minipage}%
    }
    \caption{One-loop vertex and self-energy corrections to the $Z\ell^+ \ell^-$ coupling due to heavy scalars $T^-$ and $T^{--}$. The prime does not mean derivative.}
    \label{fig:radllT}
\end{figure}
These loop functions have the same structure as those in Eq.~\eqref{eq:GammaZnunu}, up to different couplings at vertices,
\begin{align}\label{eq:GammaZellell}
&\Gamma'^{(1)} = \frac{1}{2} \Gamma^{(1)} \,,\quad \Gamma'^{(2)} = \frac{1}{2} \sin^2\theta_W  \Gamma^{(2)}\,,\quad \Sigma'^{(1)} = \Sigma^{(2)} = \frac{1}{2} \Sigma^{(1)} \,,\nn\\
&\Gamma'^{(3)} = (-1+2\sin^2\theta_W)  \Gamma^{(1)} \,,\quad
\Gamma'^{(4)} = (-1+2\sin^2\theta_W) \Gamma^{(2)}\,,\quad \Sigma'^{(2)} = \Sigma^{(1)} \ .
\end{align}

The resulting shift to the electroweak form factor is 
\begin{align}\label{eq:deltarhoellfromT}
\left(\sqrt{\rho_Z^{\ell_\alpha}} -1 \right)_{T,\, \theta=0} = - \left[ \sum_{i=1}^4 \Gamma'^{(i)} + (-1+2\sin^2\theta_W)\sum_{j=1}^2 \frac{\partial \Sigma'^{(j)}}{\partial \cancel{p}}  \right] \ .
\end{align}
Again, the result is UV finite. 
The shift in $\sqrt{\rho_Z^{\ell_\alpha}}$ and $\kappa_Z^{\ell_\alpha}$ affect both the asymmetry parameters and the $Z$-boson partial decay width, and in turn other electroweak observables, as discussed in section~\ref{sec:Zpole}.
Both Eqs.~\eqref{eq:deltagZfromT} and \eqref{eq:deltarhoellfromT} are contributions from heavy $T$ to $Z$-pole physics, thus they are evaluated at momentum transfer $q^2=m_Z^2$.

Finally, we calculate the correction to the $W\ell\bar\nu$ coupling, $g_W$ defined in Eq.~\eqref{eq:EWL}. The vertex Feynman diagrams are shown in Fig.~\ref{fig:radnuellT}. We do not show the self-energy diagrams which are identical to those in Figs.~\ref{fig:radnunuT} and \ref{fig:radllT}.
\begin{figure}[h]
    \centering
    \begin{align}
&\begin{tikzpicture}
\begin{feynman}
\vertex [label=above:\(W^-\)] (a) at (0,0);
\vertex (b) at (0.7,0);
\vertex (c) at (1.57,0.5);
\vertex (d) at (1.57,-0.5);
\vertex [label=right:\(\ell^-_\alpha\)] (e) at (2.27,0.5);
\vertex [label=right:\(\nu_\alpha\)] (f) at (2.27,-0.5);
\feynmandiagram [inline=(a.base)] {
(a)  -- [photon, momentum'=\(q\)] (b),
(b) -- [fermion, edge label=\(\nu^c\)] (c),
(d) -- [charged scalar, edge label'=\(T^-\)] (c),
(d) -- [fermion, edge label=\(\ell^c\)] (b),
(c) -- [fermion] (e),
(d) -- [anti fermion] (f),
};
\end{feynman}
\end{tikzpicture} 
\hspace{0.5cm}
\raisebox{0.65cm}{
$\equiv -i g_{W}^0 \Gamma''^{(1)}(q^2)$ \ ,
}  \nonumber \\
&\begin{tikzpicture}
\begin{feynman}
\vertex [label=above:\(W^-\)] (a) at (0,0);
\vertex (b) at (0.7,0);
\vertex (c) at (1.57,0.5);
\vertex (d) at (1.57,-0.5);
\vertex [label=right:\(\ell^-_\alpha\)] (e) at (2.27,0.5);
\vertex [label=right:\(\nu_\alpha\)] (f) at (2.27,-0.5);
\feynmandiagram [inline=(a.base)] {
(a)  -- [photon, momentum'=\(q\)] (b),
(b) -- [charged scalar, edge label=\(T^-\)] (c),
(d) -- [fermion, edge label'=\(\nu^c\)] (c),
(d) -- [charged scalar, edge label=\(T^0\)] (b),
(c) -- [fermion] (e),
(d) -- [anti fermion] (f),
};
\end{feynman}
\end{tikzpicture} 
\hspace{0.5cm}
\raisebox{0.65cm}{
$\equiv -i g_{W}^0 \Gamma''^{(2)}(q^2)$ \ ,
} 
\hspace{1cm} 
\begin{tikzpicture}
\begin{feynman}
\vertex [label=above:\(W^-\)] (a) at (0,0);
\vertex (b) at (0.7,0);
\vertex (c) at (1.57,0.5);
\vertex (d) at (1.57,-0.5);
\vertex [label=right:\(\ell^-_\alpha\)] (e) at (2.27,0.5);
\vertex [label=right:\(\nu_\alpha\)] (f) at (2.27,-0.5);
\feynmandiagram [inline=(a.base)] {
(a)  -- [photon, momentum'=\(q\)] (b),
(b) -- [charged scalar, edge label=\(T^{--}\)] (c),
(d) -- [fermion, edge label'=\(\ell^c\)] (c),
(d) -- [charged scalar, edge label=\(T^-\)] (b),
(c) -- [fermion] (e),
(d) -- [anti fermion] (f),
};
\end{feynman}
\end{tikzpicture} 
\hspace{0.5cm}
\raisebox{0.65cm}{
$\equiv -i g_{W}^0 \Gamma''^{(3)}(q^2)$ \ ,
} 
\nonumber
\end{align} 
\caption{One-loop vertex and self-energy corrections to the $W\ell \bar{\nu}$ vertex due to the heavy scalars.}
\label{fig:radnuellT}
\end{figure}
The loop functions are equal to
\begin{align}
\Gamma''^{(1)} = \frac{1}{2} \Gamma^{(1)}, \quad  
\Gamma''^{(2)} = \Gamma''^{(3)} = \frac{1}{2} \Gamma^{(2)} \ .
\end{align}
The finite correction to $g_W$ is given by
\begin{eqnarray}\label{eq:deltagWfromT}
\left.\frac{(\delta g_{W})_{\alpha\alpha}}{g_{W}^0}\right|_{T,\, \theta=0} = \sum_{i=1}^3 \Gamma''^{(i)} + \frac{1}{2} \sum_{j=1}^2\frac{\partial \Sigma^{(j)}}{\partial \cancel{p}} + \frac{1}{2} \sum_{j=1}^2\frac{\partial \Sigma'^{(j)}}{\partial \cancel{p}} \ .
\end{eqnarray}

The radiative correction to $W\ell\bar\nu$ coupling from $T$ depends on the momentum transfer of a specific charged-current process through the ratio $q^2/M^2$. It vanishes in the $M\to \infty$ limit as required by the decoupling theorem.
The correction is suppressed by $q^2$ for low-energy processes such as weak decays.~\footnote{The same is true for the heavy $T$ contributions to $\delta g_W$, $\delta g_Z$, and the $Z\ell^+\ell^-$ couplings.}  For electroweak observables, its effect is strongest in $W$-boson decay where $q^2=m_W^2$. 
We include the heavy $T$ contribution as part of $\delta g_W$ in Eq.~\eqref{eq:shiftW2body}.
In contrast, the effect is negligible in muon decay and the corresponding $\Delta r$ parameter because $q^2<m_\mu^2$. In the UV complete model, the contribution to $\Delta r$ remains the same as Eq.~\eqref{eq:defDeltar}.

In section~\ref{sec:structureofRC}, we have discussed the parametrical dependence of radiative corrections to $\delta g_V$ ($V=Z,W$) from the light $\phi$ and heavy $T$ contributions. 
Now with the concrete calculations, we are able to make a more precise comparison and highlight the interplay of  parameters in the UV complete model.

\begin{figure}[t]
    \centering
    \includegraphics[width=0.48\linewidth]{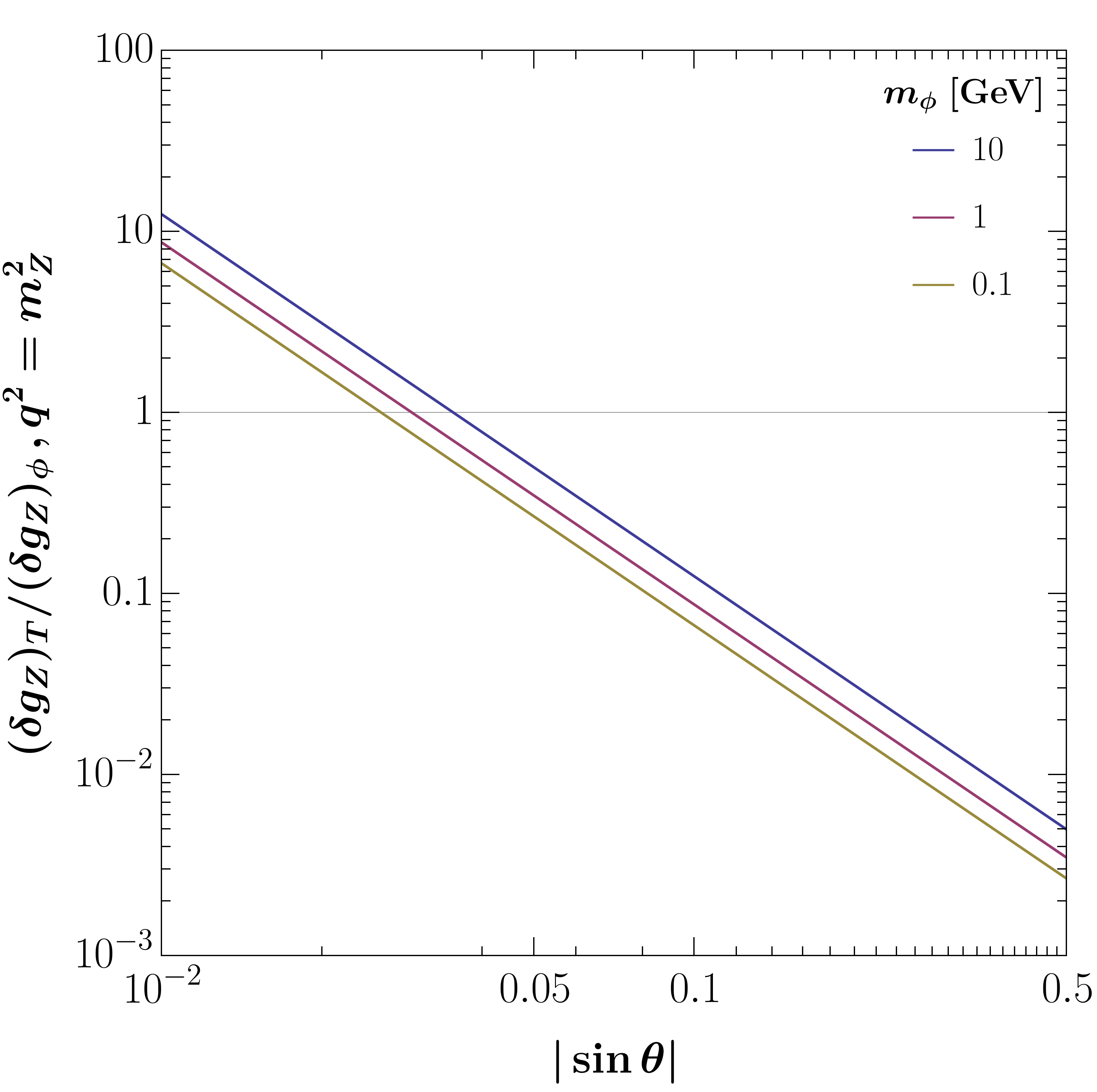}
    \includegraphics[width=0.48\linewidth]{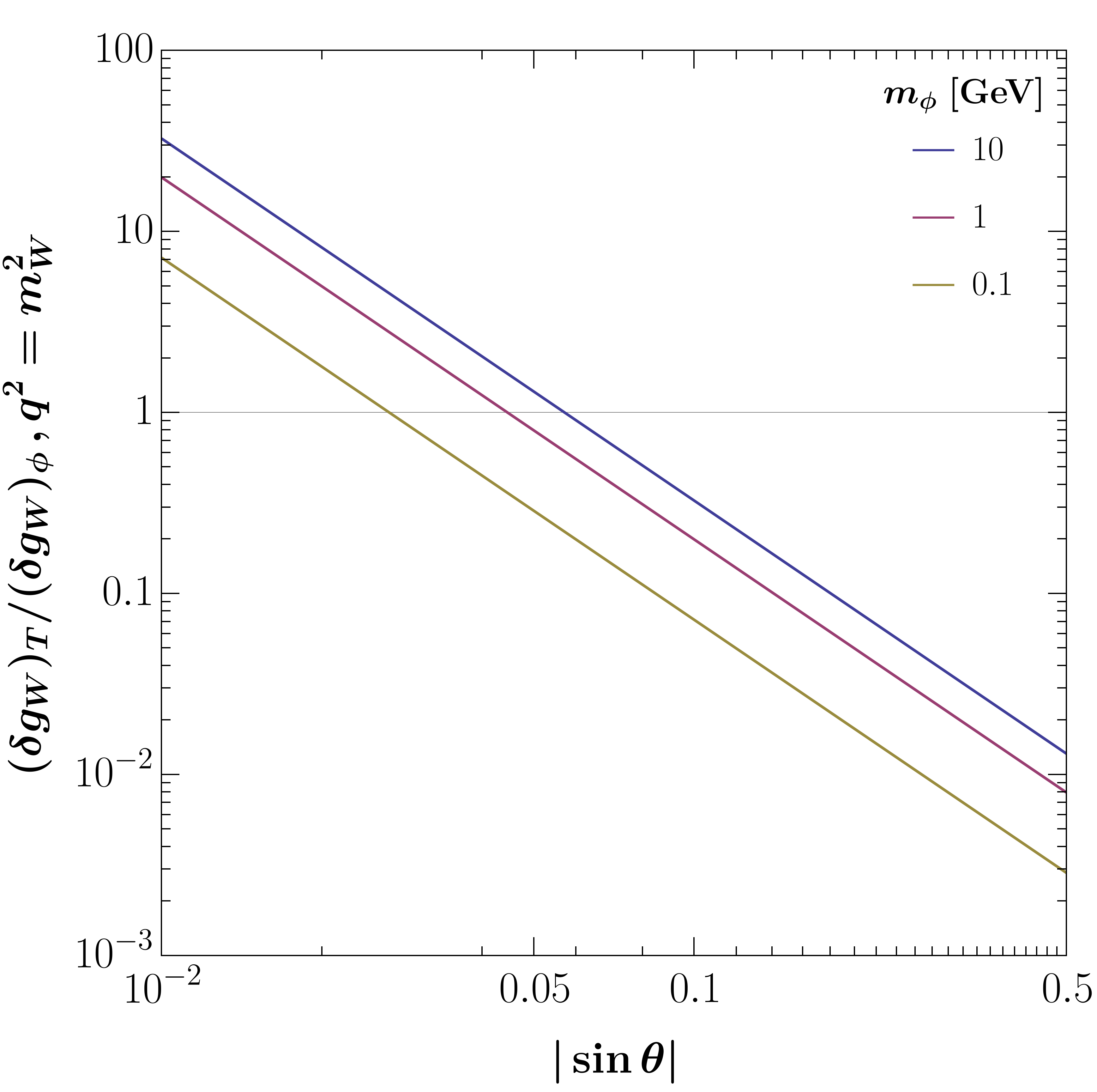}
    \caption{Ratio of the heavy $T$ versus light $\phi$ contribution to $\delta g_{Z, W}$, with respect to $\sin\theta$ for three different choice of $m_\phi$, and $M=1$ TeV.
    The heavy $T$ contribution is negligible for $\sin\theta\gtrsim 0.05$.}
    \label{fig:sinlam}
\end{figure}

The above $T$ induced radiative corrections are all proportional to the fundamental Yukawa coupling $y^2$.
Using Eq.~\eqref{eq:lambday}, for fixed $\lambda$, $y = \lambda/\sin\theta$ is inversely proportional to the mixing parameter $\sin\theta$. 
As a result, the $T$ contribution can be subdominant to that of $\phi$ for sufficiently large $\sin\theta$.
In Fig.~\ref{fig:sinlam}, we show the ratio of heavy $T$ contribution to $\delta g_{Z,W}$ derived in Eqs.~\eqref{eq:deltagZfromT} and \eqref{eq:deltagWfromT} to the light $\phi$ contribution given in Eq.~\eqref{eq:dgzdgw} as a function of $\sin\theta$ for several choices of $m_\phi$.
We hold $M=1\,$TeV fixed. The coupling $\lambda$ dependence cancels away in this ratio.
We find that the ratio is smaller than 1 for $|\sin\theta|\gtrsim 0.05$.
This is the parameter space where the light $\phi$ contribution dominates the radiative corrections and the precision electroweak constraints shown in Fig.~\ref{fig:res} continue to hold in the UV complete model.
We also note that for light $\phi$ below half of the $Z$ mass, $|\sin\theta|$ cannot be made too large due to the $Z\to\phi\phi^*$ constraint discussed in section~\ref{sec:Zphiphi*}.
The value of $|\sin\theta|$ is further bounded from above due to the unitarity constraint on $|y'|$.
Using Eq.~\eqref{eq:thetasmall}, setting $M=1\,$TeV and requiring $|y'|<4\pi$, we obtain 
\begin{equation}\label{eq:pert}
|\sin\theta|< 0.38 \ .
\end{equation}

In the above results, Eqs.~\eqref{eq:deltagZfromT}, \eqref{eq:deltarhoellfromT} and \eqref{eq:deltagWfromT}, a small mass splitting among the triplet components, $\Delta M$, has been neglected. This amounts to neglecting terms suppressed by $\Delta M/M \ll1$. The mass splitting plays a more important role in the contribution to electroweak oblique parameters, as will be discussed in section~\ref{sec:oblique}.

\subsection{Constraints from Global Fit}\label{sec:UVglobalfit}

Here we discuss the procedure of precision electroweak fit in the UV complete model with the $SU(2)_L$ scalar triplet. Compared to the simplified model explored in section~\ref{sec:globalfit}, the UV completion has two additional parameters. The complete set of beyond SM parameters are
\begin{align}
    \{\lambda, m_\phi, \sin\theta, M, \Delta M\} \ .
\end{align}

\begin{figure}[t]
    \centering
    \includegraphics[width=0.48\linewidth]{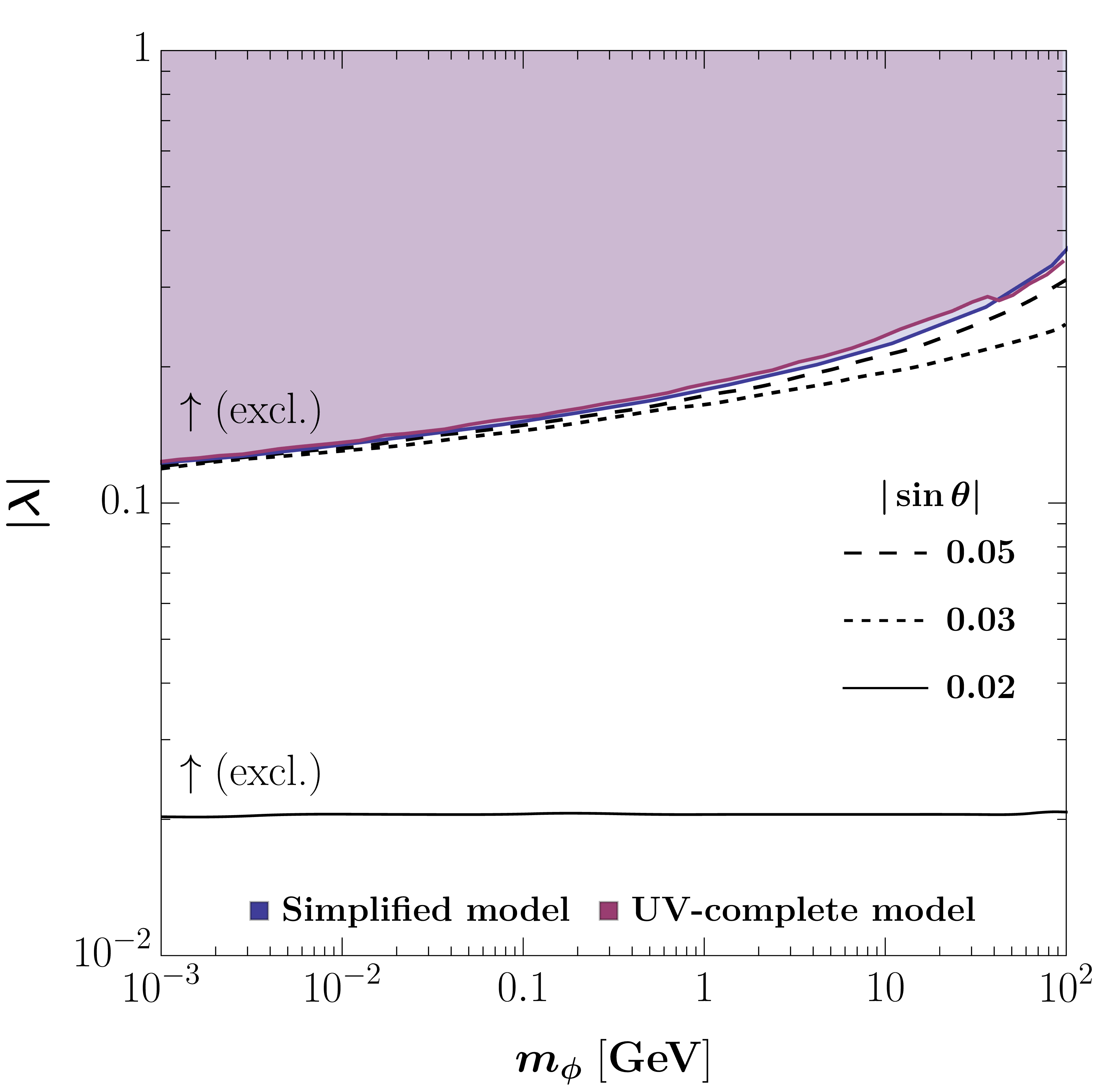}
    \includegraphics[width=0.48\linewidth]{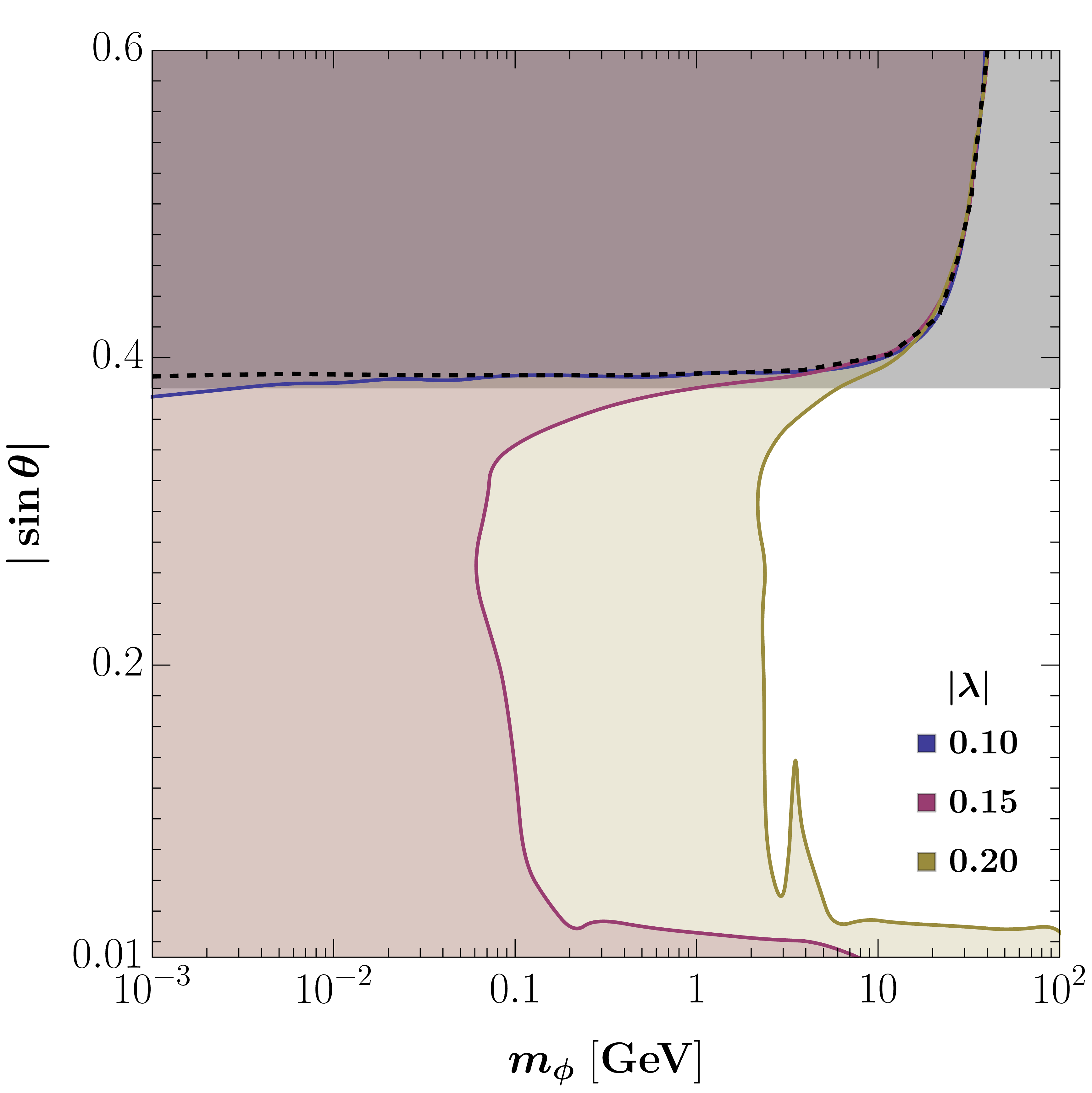}
    \caption{Constraints on the UV-complete model for the flavor-universal case. The shaded regions are excluded at the $2\sigma$ level. \textit{Left:} The dashed, dotted and solid black curves correspond to fixed values of $|\sin\theta|$, while purple and blue shaded regions represent the results of the global analysis for the UV-complete model ($|\sin\theta|\in \{0.01,1.0\}$) and the simplified model, respectively. \textit{Right:} Exclusion curves in the $|\sin\theta|$ vs. $m_\phi$ plane for fixed values of $\lambda$. The dashed curve denotes the exclusion from the global fit where $|\lambda| \in \{0.01-1.0\}$. The gray region is the theoretical upper bound on $|\sin\theta|$ from Eq.~\eqref{eq:pert}.}
    \label{fig:uvres}
\end{figure}

To reduce number of parameters in the scanning, we make two simplifications to our analysis. First, the LHC experiment already constrains the doubly-charged scalar mass to be above $\sim900$\,GeV~\cite{ATLAS:2022pbd, Guedes:2025jqu, Banerjee:2025qkb}. 
Meanwhile, the precision electroweak fit is sensitive to $\lambda, \sin\theta \sim \mathcal{O}(0.1)$, which corresponds to $M$ not far above TeV scale, through Eqs.~\eqref{eq:thetasmall} and \eqref{eq:lambday}.
Thus we hold $M=1\,$TeV fixed in the analysis.  
Furthermore, as will be discussed in section~\ref{sec:oblique}, the mass splitting parameter $\Delta M$ mainly controls the oblique parameter $T$ and is strongly correlated to $m_\phi$ and $\sin\theta$. 
We do not include the heavy triplet contribution to $S,T,U$ in the global fit. Instead, their ranges are determined posteriorly from the fit, which is then used to solve for $\Delta M$ consistent with the oblique constraints.
These considerations help to reduce the free parameters of scanning down to three
\begin{align}
    \{\lambda, m_\phi, \sin\theta\} \ .
\end{align}
The parameter ranges of our scan are 
\begin{align}
\lambda \in (10^{-2}, 1), \quad \sin\theta \in(10^{-2}, 1), \quad m_\phi \in (10^{-3}, 10^2)\, {\rm GeV} \ .
\end{align}
The results are shown in Fig.~\ref{fig:uvres}.
We focus on case where $T$ and $\phi$ couple universally to the three neutrino flavors. Similar results hold for flavor specific cases.
In the left panel, the purple shaded region is excluded at $2\sigma$ level in the $\lambda$ versus $m_\phi$ parameter space where we allow $\sin\theta$ to float in the scan. This corresponds to the most conservative limit in the UV complete model. The best-fit point corresponds to $\sin\theta=0.26$, with $\chi^2_{\rm min}= 13.22$. The corresponding $p$-value $P=67\%$ is determined by the toy MC analysis.
In the same plot, the blue shaded region corresponds to the $2\sigma$ exclusion in the simplified model derived early on and shown as the blue shaded region in Fig.~\ref{fig:nsi} (left). 
The very small difference between the purple and blue shaded regions implies that resorting to the UV completion does not significantly improve the global fit to the electroweak observables.
Near the best fit point with $|\sin\theta|=0.26$, it is good enough to use the simplified model to set constraint on $\lambda$ versus $m_\phi$.

Because the value of $\sin\theta$ is set by nature rather than by the authors, we also perform scans in the $\lambda$ versus $m_\phi$ plane for several other choices of $\sin\theta$ equal to 0.05, 0.03, 0.02. The results are shown as the black solid, dashed and dotted lines, respectively.
We find the upper bound on $\lambda$ gets stronger and less $m_\phi$ dependent for smaller values of $\sin\theta$.
This behavior is consistent with the discussions around Fig.~\ref{fig:sinlam}.
For very small $\sin\theta$, the heavy $T$ contribution to radiative corrections is enhanced and dominates over that from light $\phi$. 
On the other hand, for $\sin\theta\gtrsim 0.03$, $\phi$ dominates and the upper bound on $\lambda$ is very similar to the simplified model result.

In the right panel of Fig.~\ref{fig:uvres}, the black dashed curve shows the precision electroweak constraint, upper bound on $|\sin\theta|$ as a function of $m_\phi$, where we allow $|\lambda|$ to float between 0.01 and 1.
We also use the shaded regions to show the exclusion for several cases of fixed $|\lambda|$ values. All are at $2\sigma$ level. 
The parameter space with $m_\phi< m_Z/2$ and $|\sin\theta| \gtrsim 0.4$ is experimentally excluded due to too large $Z\to \phi\phi^*$ decay width, as discussed in section~\ref{sec:Zphiphi*}.
For intermediate range of the mixing parameter $0.03 \lesssim |\sin\theta|\lesssim 0.4$, a larger $|\lambda|$ leads to a stronger lower bound on $m_\phi$, which is consistent with the shape of exclusion curve in the left panel. 
For $|\lambda|=0.15$ and $0.2$, the very small $|\sin\theta|$ region is excluded regardless of $m_\phi$ because of the heavy $T$ dominance. 
Similar floor also exists for smaller $|\lambda|=0.1$ but only excludes low $|\sin\theta|$ values that lie outside the plot range. 

\subsection{Contribution to Electroweak Oblique Parameters}\label{sec:oblique}

Finally, we discuss the loop corrections from the triplet $T$ through the vacuum polarization diagrams of the SM vector bosons. 
At energy scales higher than $M$, the triplet contributes to the beta functions of $SU(2)_L$ and $U(1)_Y$ gauge couplings. This does not bring additional constraint, due to the lack of precision measurements at those high energies.
For momentum transfers well below the triplet mass, their remnant effects in the gauge boson self-energy diagrams are accounted for by the electroweak oblique parameters~\cite{Peskin:1990zt, Peskin:1991sw, Maksymyk:1993zm}. 

The oblique parameters contributed by a $SU(2)_L$ triplet scalar have been explored in the literature~\cite{Chun:2012jw, Mandal:2022zmy, Cheng:2022hbo}. Here, we generalize those results by introducing a light scalar $\phi$ and switching on its mixing with $T^0$. 
Following the notations used in~\cite{Lavoura:1993nq},
\begin{equation}
\begin{split}
S=& \frac{16\pi c^2 s^2}{e^2} \left[ \frac{A_{ZZ}(M_Z) - A_{ZZ}(0)}{M_Z} - \frac{c^2 - s^2}{cs} \left.\frac{\partial A_{\gamma Z}(P)}{\partial P}\right|_{P=0} - \left.\frac{\partial A_{\gamma \gamma}(P)}{\partial P}\right|_{P=0} \right] \ , \\
T=& \frac{4\pi}{e^2} \left[ \frac{A_{WW}(0)}{M_W} - \frac{A_{ZZ}(0)}{M_Z} \right] \ , \\
U=&\frac{16\pi s^2}{e^2} \Bigg[ \frac{A_{WW}(M_W) - A_{WW}(0)}{M_W} - c^2 \frac{A_{ZZ}(M_Z) - A_{ZZ}(0)}{M_Z} - 2cs \left.\frac{\partial A_{\gamma Z}(P)}{\partial P}\right|_{P=0}\\
&- s^2 \left.\frac{\partial A_{\gamma \gamma}(P)}{\partial P}\right|_{P=0} \Bigg] \ ,
\end{split}
\end{equation}
where $s = \sin\theta_W$, and $c=\sqrt{1-s^2}$. Here, the capital $M_i$'s represent the corresponding mass square, {\it i.e.}, $M_W = m_W^2$, $M_Z = m_Z^2$. The capital
$P$ is equal to $p^2$ with $p^\mu$ being the external momentum of each vector polarization diagram. 

The vacuum polarization amplitudes are given by
\begin{align}\label{eq:VP}
A_{\gamma\gamma} (P)=& \frac{e^2}{8\pi^2} \left[ 4 F(M_{-1}, M_{-1}, P) + F(M_0, M_0, P) \right]\,,\nn \\
A_{\gamma Z} (P)=& \frac{e^2}{8\pi^2cs} \left[  2 (c^2-s^2) F(M_{-1}, M_{-1}, P)  - s^2 F(M_0, M_0, P) \right]\,,\nn\\
A_{WW}(P)=& \frac{e^2}{8\pi s^2} \left\{ F(M_{0}, M_{-1}, P) + \left[ \cos^2\theta F_s(M_{+1}) + \sin^2\theta F_s(M_\phi) \right] + F_s(M_0)  \right.\nn \\
&+ \left.   \left[ \cos^2\theta F_b(M_{+1}, M_0, P) + \sin^2\theta F_b(M_\phi, M_0, P) \right]\right\}\,,\nn \\
A_{ZZ}(P)=& \frac{e^2}{8\pi^2c^2s^2} \left\{  (c^2-s^2)^2 F(M_{-1}, M_{-1}, P)  + s^4 F(M_0, M_0, P) + 2 \cos^2\theta F_s(M_{+1})  \right. \nn\\
&+ 2 \sin^2\theta F_s(M_\phi)+ \left. \left[ \cos^4\theta F_b(M_{+1}, M_{+1}, P)  + 2 \sin^2\theta \cos^2\theta F_b(M_{+1}, M_\phi, P)\right. \right.\nn\\
&+\left.\left. \sin^4\theta F_b(M_\phi, M_\phi, P) \right] \right\}\,,
\end{align}
where $M_{+1}, M_0, M_{-1}$ denote the mass squares of the triplet components $T^0, T^-, T^{--}$ given in Eq.~\eqref{eq:tripletcomponentmasses}, and $\sin\theta$ is the $\phi$-$T^0$ mixing parameter introduced in Eq.~\eqref{eq:thetasmall}. 
Because $T^0$ is electric neutral, $A_{\gamma Z}$ and $A_{\gamma\gamma}$ do not depend on $M_{+1}$.

\begin{figure}[t]
    \centering
    \includegraphics[width=0.48\linewidth]{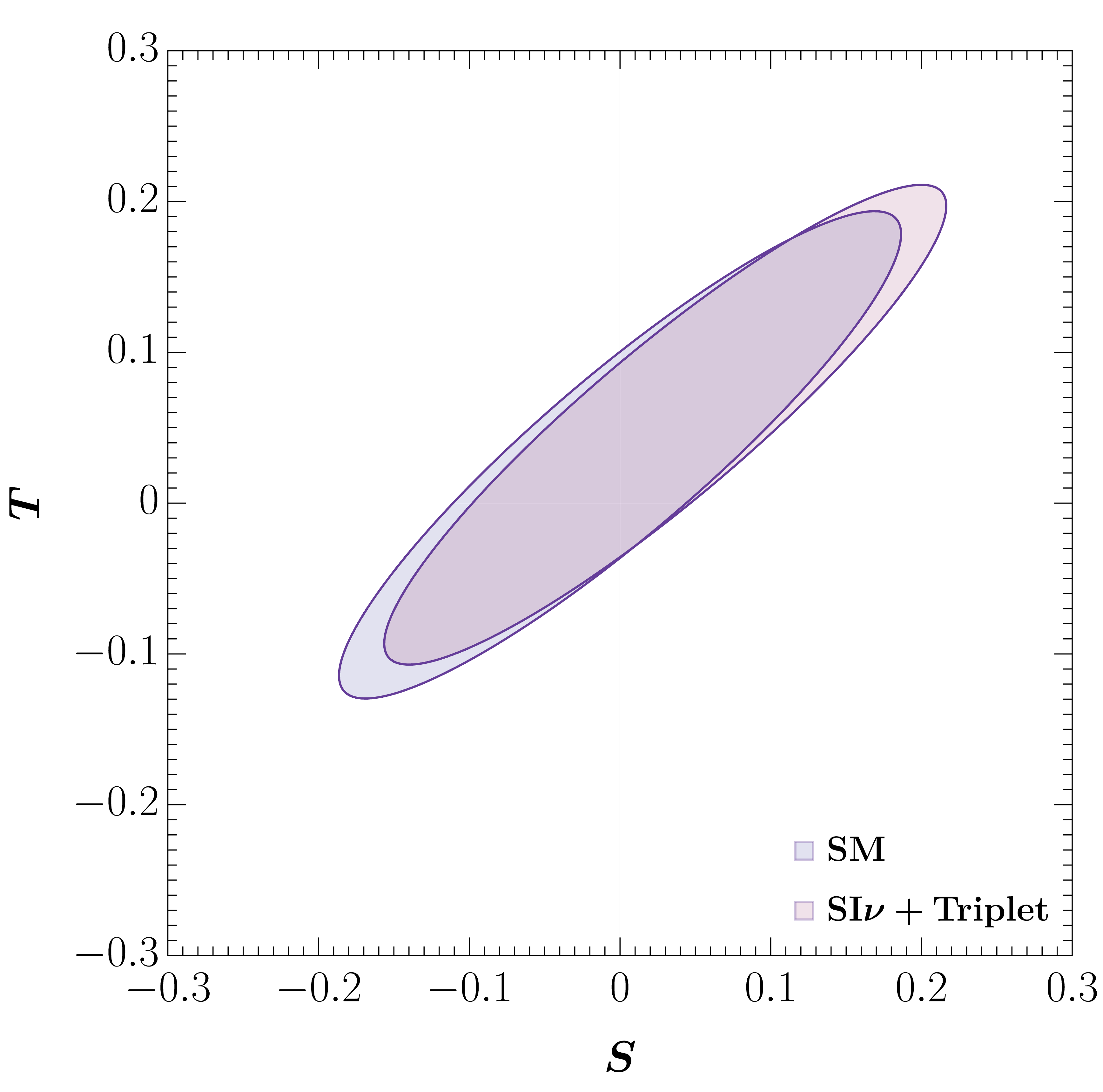}
    \includegraphics[width=0.48\linewidth]{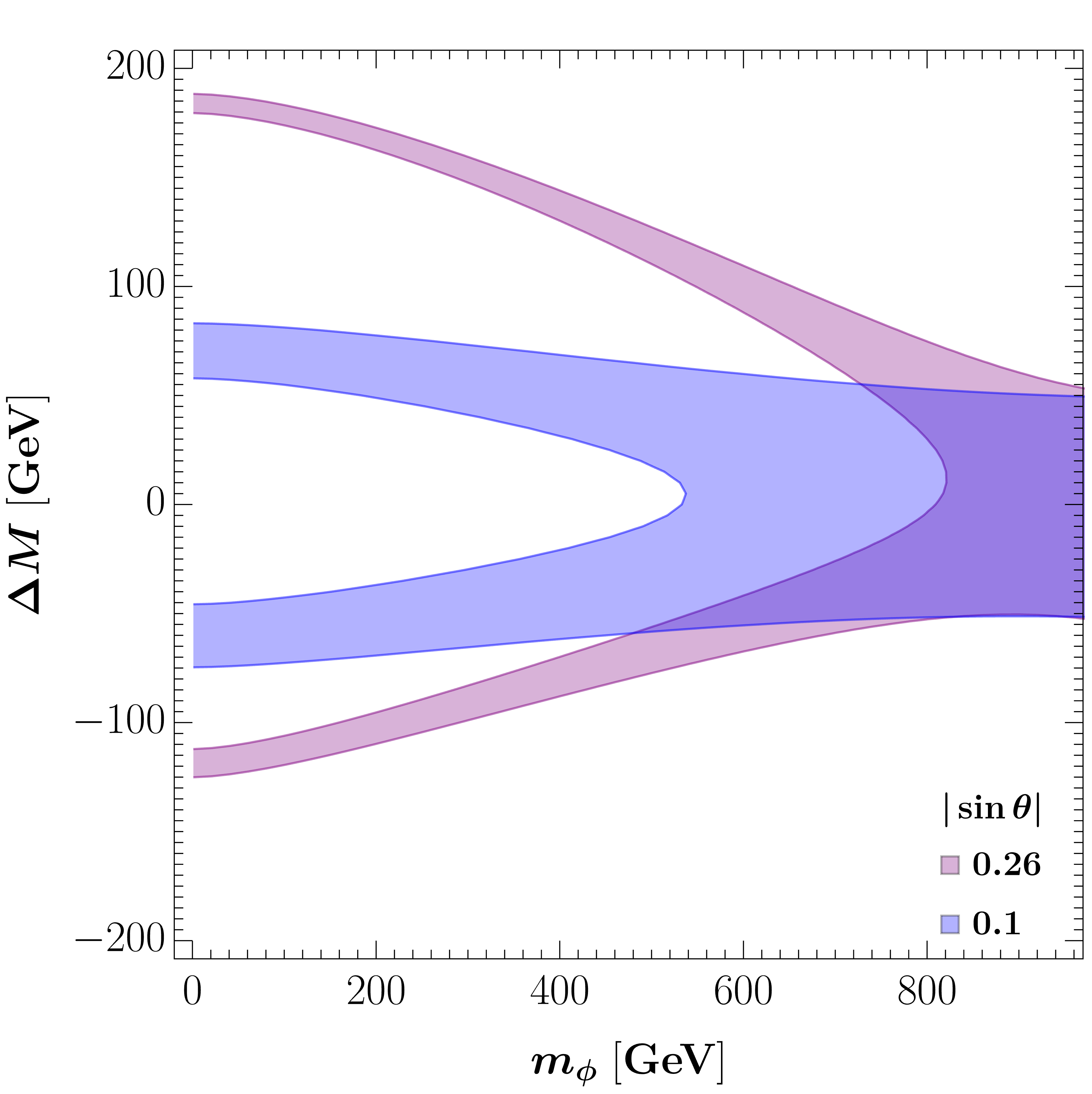}
    \caption{\textit{Left:} Electroweak precision constraints ($2\sigma$) on $S$ and $T$ ($U = 0$) in the SM and the UV complete model after including all the other corrections. \textit{Right:} Correlation between $\Delta M$ and $m_\phi$ for $M=1\,$TeV and two choices of $\sin\theta$. The shaded regions correspond to $S,T$ within the ranges, $T = 0.052\pm0.064$ and $S = 0.03\pm0.075$, consistent with the purple ellipse in the left panel.}
    \label{fig:oblique}
\end{figure}

\begin{figure}[b]
    \centering
    \feynmandiagram [horizontal=a to b, layered layout] {
  a -- [photon, momentum'=\(p\)] b [dot] -- [out=125, in=55, loop, min distance=2cm, charged scalar] b -- [photon] c,
};
 \hspace{2cm}
\raisebox{1.2cm}{
\feynmandiagram [inline=(a.base), horizontal=a to d] {
  i1] -- [photon, momentum'=\(p\)] b -- [charged scalar, half left] a -- [photon] i2],
  a -- [charged scalar, half left, looseness=1.5] b,
}; }
    \caption{Sunrise and bubble diagrams for $Z,W$ self-energy diagrams from scalar triplet.}
    \label{fig:ZZ}
\end{figure}
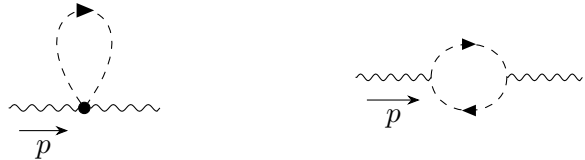

The functions $F_s, F_b$ are the finite part of the sunrise and bubble diagrams in Fig.~\ref{fig:ZZ},
\begin{align}
F_s(M) & = \frac{1}{2} M \ln M\,, \\
F_b(M_1, M_2, P) & = - \int_0^1 dx \left[ (x^2 -x) P + x M_1 + (1-x)M_2 \right] \ln \left[ (x^2 -x) P + x M_1 + (1-x)M_2 \right]\,,\nn
\end{align}
and we further write
\begin{equation}
\begin{split}
F(M_1, M_2, P) &=  F_s(M_1) + F_s(M_2) + F_b(M_1, M_2, P)  \ , \\
F(M, M, P) &= 2F_s(M) + F_b(M, M, P)  \,.
\end{split}\label{eq:F}
\end{equation}
A useful identity for our calculation is
\begin{equation}
\left.\frac{\partial F(M, M, P)}{\partial P} \right|_{P=0} = \frac{1}{6} (1 + \ln M) \ .
\end{equation}

The impact of $S,T,U$ parameters on electroweak observables are listed in appendix B of~\cite{Peskin:1991sw}. 
In the $\theta=0$ limit, the contribution from a triplet with degenerate masses at TeV scale is safely small given the existing constraints.
This is no longer the case in the presence of $T^0$ mixing with a light $\phi$, which contributes to custodial symmetry violation at loop level and can give a potentially large contribution to the $T$ parameter.

The global analysis discussed earlier has taken into account the radiative corrections to the gauge-fermion couplings calculated in section~\ref{sec:gaugevertex} but not the oblique parameters. 
Instead, the allowed range of $S, T, U$ parameters for the best fit point is obtained as a spin-off of the analysis done by $\gfit$. The result is depicted by the purple ellipse in the left panel of Fig.~\ref{fig:oblique}.
We only show the $S$ versus $T$ plane because the constraint on $U$ is much weaker.
For comparison, we use the blue ellipse to show the ranges of $S,T$ in the SM.
The difference between the two ellipses is small, which is another evidence showing that self-interacting neutrino models (simplified or UV completed) does not substantially improve the electroweak global fit over the SM.

In the right panel of Fig.~\ref{fig:oblique}, we show the $\Delta M$ versus $m_\phi$ parameter space where the predictions for $S, T$ parameters are within the purple ellipse in the left panel.
We hold $M=1\,$TeV fixed and the blue and purple regions correspond to $\sin\theta=0.1$ and 0.26, respectively. 
For small $m_\phi$, we find that a mass splitting along the triplet components $\Delta M$ of order $\pm100$\,GeV is needed.
It is mainly driven by suppressing the $T$ parameter.
Due to the strong correlation between $\Delta M$, $m_\phi$ and $\sin\theta$, we did not directly include the electroweak oblique parameters in the global fit.

As the main message from the discussions of this subsection, it is always possible to find a proper $\Delta M$ value to suppress the oblique parameters.
This requires some tuning among the UV complete model parameters, and it is the price one needs to pay in order to accommodate a light $\phi$ that mediates strong neutrino self-interaction.

%%%%%%%%%%%%%%%%%%%%%%%%%%%%
\section{Conclusions and Outlook}
\label{sec:sum}
%%%%%%%%%%%%%%%%%%%%%%%%%%%

Novel neutrino self-interaction with strength much higher than predicted by the Fermi theory continues to serve as a well-motivated candidate of new physics. 
Extension of the SM with a light neutrinophilic scalar $\phi$ has been employed as the leading model and has been the target of various probes in the laboratories and cosmology.
In this work, we perform an in-depth study of one-loop radiative corrections in such a model and derive, for the first time, precision electroweak constraints on the  self-interacting neutrino parameter space.
Our investigation is made in both simplified model with only $\phi$ and a concrete UV completion with additional neutrinophilic scalars.

In the simplified model, radiative corrections to the electroweak sector can be computed with the guidance of gauge invariance restoration and the Ward identity. The general forms of neutrino neutral- and charged-current couplings ($g_Z, g_W$) are presented in Eq.~\eqref{eq:dgzdgw}, which features a logarithmic enhancement factor in the light $\phi$ limit. The presence of the log factor does not depend on the UV completion details.
The correction to $g_Z$ mainly contributes to the invisible $Z$ decay width, whereas the correction to $g_W$ can affect the Fermi constant measurement. The latter occurs through the $\Delta r$ parameter which in turn modifies the prediction for a number of electroweak observables as detailed in section~\ref{sec:ew}. We modify the public $\gfit$ code to include all the effects and perform a global fit to the most recent data from lepton and hadron colliders.
The results are summarized in Fig.~\ref{fig:res} which represent the state-of-art constraints on the neutrinophilic coupling $\lambda$ as a function of $m_\phi$. The precision electroweak constraints derived in this work are the leading ones for $m_\phi$ above a few hundred MeV scale.
For the flavor universal and $\nu_e/\nu_\mu$ specific neutrino self-interactions, the constraints are much stronger than the previous limit derived only using the invisible $Z$ width.
This further allows us to report the highest value of $G_{\rm eff}$
\begin{equation}
   G_{\rm eff} < \left\{ \begin{array}{cl}
      1.4\times 10^{-6}\,{\rm MeV}^{-2},\, & \quad \text{flavor universal case} \\
   3.5\times 10^{-6}\,{\rm MeV}^{-2},\, & \quad \nu_e \text{ specific case} \\
   4.1\times 10^{-6}\,{\rm MeV}^{-2},\, & \quad \nu_\mu \text{ specific case} \\
   2.3\times 10^{-4}\,{\rm MeV}^{-2},\, & \quad \nu_\tau \text{ specific case}
    \end{array}\right.
\end{equation}
These findings will serve as a road map for further explorations on neutrino cosmology and origin of dark matter.
The correct relic abundance of sterile neutrino dark matter can be produced through the oscillation mechanism in the presence of neutrino self-interaction~\cite{DeGouvea:2019wpf} and our precision electroweak constraints are relevant for the 
large and intermediate $\phi$ mass regions of the relic target.
Our new constraints also facilitate interesting interplay with upcoming experimental searches.
As an example, consider the lower-right panel of Fig.~\ref{fig:res} for the case of flavor universal neutrino self-interaction. Currently, the largest $G_{\rm eff}$ is found at the intersection of PIENU and BBN constraints. If future pion decay experiments ({\it e.g.},~\cite{PIONEER:2025idw}) improve the sensitivity by another order of magnitude, the largest $G_{\rm eff}$ value will shift to the intersection of NA62 and precision electroweak constraints.

We assess the robustness of the above results by considering a concrete UV complete model which extends the SM with $\phi$ and an $SU(2)_L$ triplet scalar $T$. The mass of $T$ lies around the TeV scale and the simplified model is obtained by integrating out the triplet. 
It is straightforward to calculate the radiative corrections in such a renormalizable and gauge invariant model.
We compare the new corrections involving the heavy $T$ in the loop diagrams to those in the simplified model.
An important parameter that controls their relative importance is the mixing between $\phi$ and the electric neutral component of the triplet $T^0$, which we call $\sin\theta$.
For $\sin\theta \gtrsim 0.03$, we find that the light $\phi$ contribution dominates and the upper bound on $\lambda$ shown in Fig.~\ref{fig:res} remain intact. 
However, for sufficiently small values of $\sin\theta$, the triplet becomes strongly coupled and its contribution can overwhelm that of $\phi$, resulting in much stronger bounds as shown in Fig.~\ref{fig:uvres}.
Another effect of the $\phi$-$T^0$ mixing is the potential large contribution to the electroweak oblique parameter $T$.
We point out that it can be suppressed with a suitable choice of the triplet mass splitting parameter $\Delta M$.
These moving parts decides whether the constraints found in the simplified model apply and when the additional contributions in the UV complete model must be taken into account.

We end with a brief comment on another possible UV completion which resorts to heavy vectorlike neutrino $N$ instead of the scalar triplet. The light scalar $\phi$ originally couples to $N$. Its neutrinophilic coupling $\lambda$ in the simplified model can be generated by the mixing between $N$ and light neutrinos.
By construction, $\lambda$ contains two powers of the $\nu$-$N$ mixing, in contrast to the single power of $\phi$-$T^0$ mixing in the scalar triplet model.
As a result, we expect stronger constraints on $\lambda$ in this alternative UV completion.
A detailed exploration of it is beyond the scope of the present work and will be done elsewhere.

\acknowledgments

We thank Ayres Freitas, Douglas Tuckler and Antonio Coutinho for useful discussions, and Ayres Freitas for commenting on the draft. 
DV acknowledges support from the Canada First Research Excellence Fund through the Arthur B. McDonald Canadian Astroparticle Physics Research Institute.
This work is supported by Subatomic Physics Discovery Grants (individual) from the Natural Sciences and Engineering Research Council of Canada.
 
\bibliographystyle{JHEP}
\bibliography{References}

\end{document}